\documentclass[aip, jcp, amsmath, amssymb, preprint]{revtex4-2}

\usepackage{mathtools}
\usepackage[dvipsnames]{xcolor}
\usepackage{tabularx}
\usepackage{dcolumn}
\usepackage{comment}
\usepackage[normalem]{ulem}
\usepackage{graphicx}
\usepackage{booktabs}
\usepackage[caption=false]{subfig}
\begin{document}

\title{A Density Functional Tight Binding Model for Rapid Screening of Spin-Orbit-Coupling-Driven Topological Properties in Bismuth}

\author{Nir Goldman}
\affiliation{Physical and Life Sciences Directorate,
             Lawrence Livermore National Laboratory, Livermore, CA 94550 USA}
\email{ngoldman@llnl.gov}
\affiliation{Department of Chemical Engineering, University of California, Davis, California 95616, United States}

\author{Ben Hourahine}
\affiliation{SUPA, Department of Physics, University of Strathclyde, Glasgow G4 0NG, UK}

\date{\today}

\begin{abstract}

We present a transferable Density Functional Tight Binding (DFTB) model for bismuth with explicit spin-orbit coupling that enables exploration of the structural degrees of freedom governing topological properties with minimal computational resources. We have used our DFTB model to run high-throughput calculations at a scale that would be significantly more expensive with standard Kohn-Sham Density Functional Theory (DFT) to determine a strain-buckling map of electronic and topological properties for bismuthene, where we identify regions of band gap closure and band inversion as a function of structural parameters. We show that our approach yields an accurate band structure for the rhombohedral phase, where small band gaps exist at or near time-reversal-invariant momentum (TRIM) points. We then demonstrate the transferability of our model to bismuth properties not included in our training data through determination of the rhombohedral phase phonon density of states and energetic ordering of additional bismuth solid phases. Finally, we have validated our topological predictions through comparison to DFT results for the optimized bilayer geometries, a band inversion order parameter, and the $\mathbb{Z}_2$ invariant as a function of monolayer unit cell length. Our efforts establish DFTB as a rigorous approach for topological characterization while yielding approximately three orders of magnitude increase in computational efficiency relative to DFT. Our model can thus serve as a computational tool for topological materials discovery by enabling large-scale computational screening to address the potentially complex relationship between physical-chemical properties and the resulting topological classification.

\end{abstract}

\maketitle

\section{Introduction}

\begin{figure}[!htp]
    \centering
    \subfloat[Rhombohedral bismuth primitive cell.\label{fig:rhomb_structure}]{
        \includegraphics[
            width=0.5\columnwidth
        ]{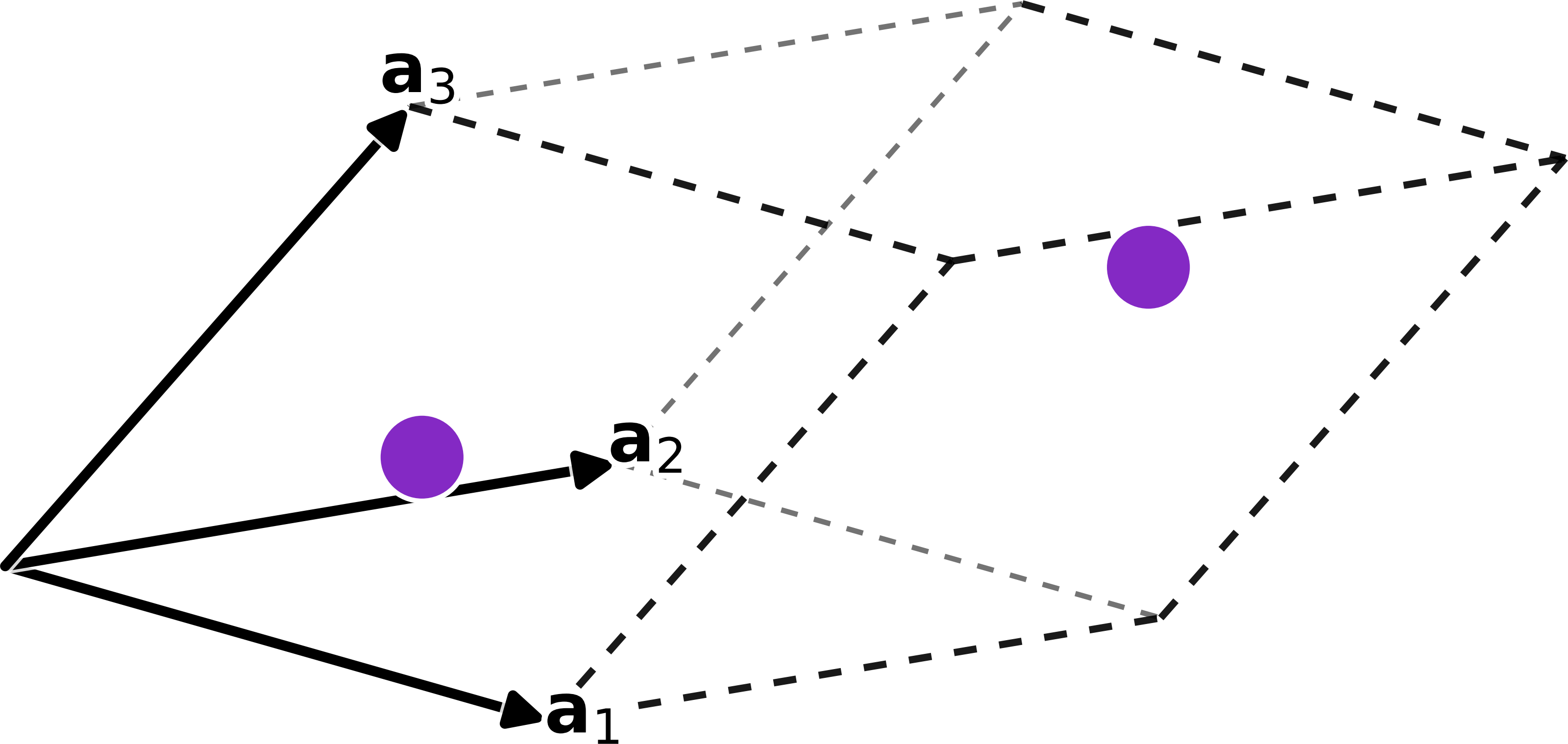}
    }
    \vspace{2mm}
    \subfloat[Buckled bismuthene structure, with example $a$ and $\Delta z$ structural parameters. Atoms are color-coded according to the layer in which they reside.\label{fig:bismuthene_structure}]{
        \includegraphics[
            width=0.9\columnwidth
        ]{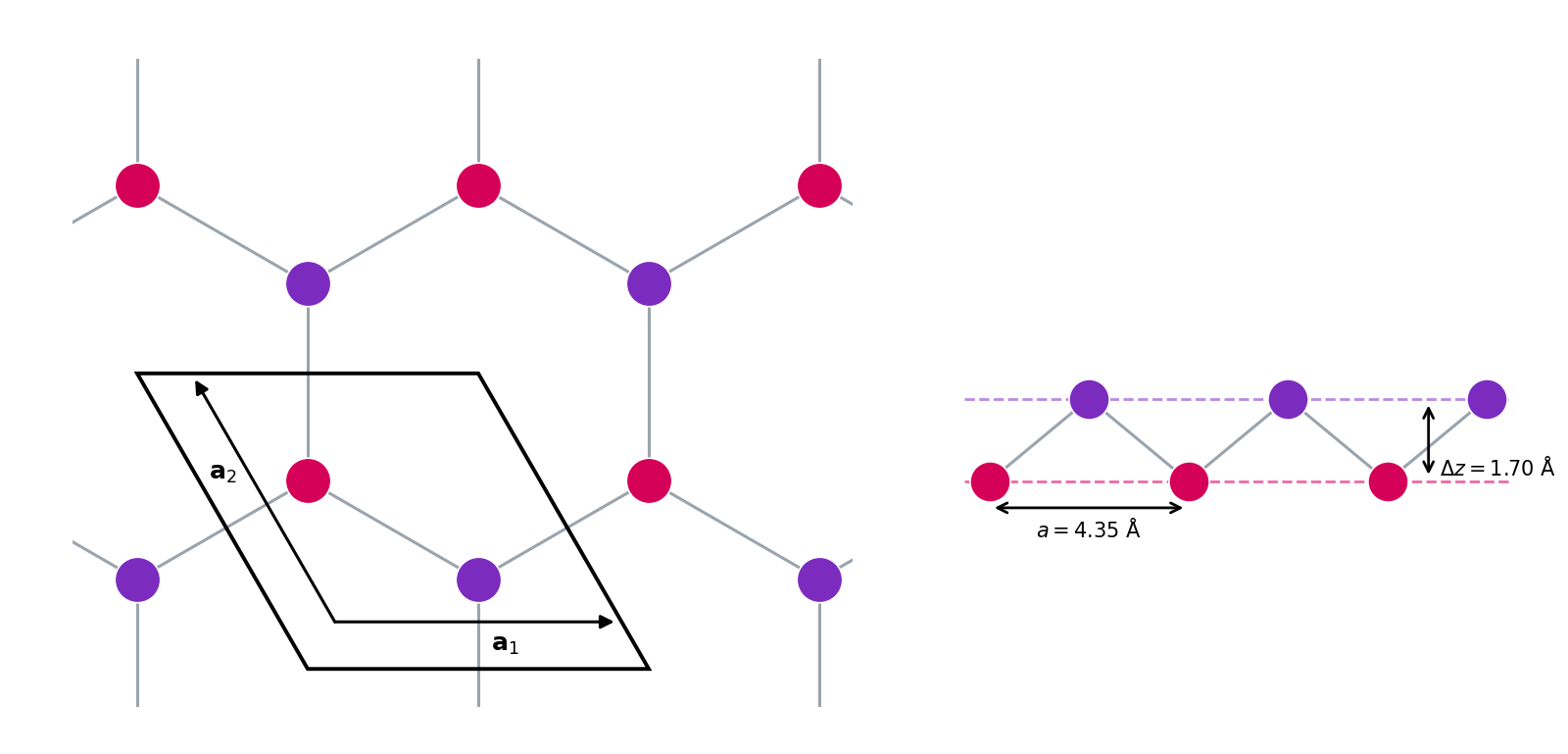}
    }
    \caption{Bismuth configurations used as training data in our study.}
    \label{fig:structures}
\end{figure}

Spin-orbit coupling (SOC) induced topological properties of bismuth-containing materials remains an active research topic due to the delicate interplay between material structure and the underlying electronic properties.\cite{Bi_band_structure} The semimetallic rhombohedral ground state (Figure~\ref{fig:rhomb_structure}) has been shown to be a higher-order topological insulator in a joint theoretical and experimental study, where the outer edges of the crystal host topologically protected conducting ``hinge'' modes.\cite{Bi_HOTI} Bismuthene (a two-dimensional allotrope with a honeycomb lattice, analogous to graphene) has received considerable attention due to its quantum spin Hall behavior while retaining a large band gap.\cite{Murakami2006} Buckled bismuthene (also called a Bi(111) bilayer; Figure~\ref{fig:bismuthene_structure}) has been shown theoretically to contain conducting edge states that spread throughout the Brillouin zone.\cite{Wada2011} Computational studies through Density Functional Theory (DFT) and empirical tight binding calculations have investigated the effects of strain,\cite{Huang2013, Huang2014} electric fields,\cite{Chen2013} and substrates\cite{Bieniek2017} on bismuthene to determine regions of trivial versus non-trivial topology as well as insulating versus metallic phases. Additional experimental studies of Bi bilayers on bulk bismuth\cite{Drozdov2014} and silicon carbide\cite{Reis2017} substrates confirm the presence of large band gaps and conductive edge states. However, a complete investigation of the bismuthene topological phase space is somewhat limited by the computational expense of DFT calculations, which can preclude high-throughput computational explorations. In addition, empirical tight binding is generally used as an analytical tool rather than in a predictive capacity for materials exploration. Future experiments would benefit from a computationally inexpensive approach that could harness the predictive accuracy and generality of DFT electronic structure calculations at a fraction of the cost.

In this regard, Density Functional Tight Binding (DFTB) is a semi-empirical quantum method derived directly from Kohn-Sham DFT that has proven to be an effective DFT-alternative for systems ranging from small molecules,\cite{Elstner_DFTB_GPR018} to liquids,\cite{Goyal_DFTB3_water_2014} heterogeneous systems,\cite{Irle2013, Irle2020, Irle2022} and high-Z solids.\cite{Hourahine10, Aradi17, Goldman_DFTB_H_Pu, Goldman_PuO2H_2022} A complete discussion of the DFTB formalism with self-consistent charges (SCC) is included elsewhere\cite{DFTB+_scc,Koskinen09,DFTB3,DFTB+,DFTB+_current,  DFTB+_2025}. Briefly, DFTB assumes neutral, spherically symmetric, atom-centered charge densities and performs a second-order Taylor expansion of the Kohn-Sham total energy in charge fluctuations. This results in the following total energy expression:

\begin{equation}
E_{\rm{DFTB}} = E_{\rm{BS}} + E_{\rm{Coul}} + E_{\rm{Rep}}.
\end{equation}

\noindent Here, $E_{\rm{BS}}$ corresponds to the band structure energy, which is determined from the approximate Kohn-Sham eigenstates using a minimal atomic basis set, and $E_{\rm{Coul}}$ is the charge fluctuation term. Both $E_{\rm{BS}}$ and $E_{\rm{Coul}}$ are computed self-consistently. Finally, $E_{\rm{Rep}}$ is the repulsive energy, which corresponds to ion-ion repulsion as well as Hartree and exchange-correlation double-counting terms. $E_{\rm{Rep}}$ is commonly taken to be short-ranged (e.g., bonding or nearest neighbor distances) and is determined via empirical function optimization through comparison to high-level quantum or experimental reference data. DFTB calculations thus provide an advantage over standard empirical tight binding by solving an approximate self-consistent Kohn-Sham equation directly, rather than treating the Hamiltonian matrix elements as empirically parameterized functions.

In this work, we detail the creation of a general DFTB approach for bismuth in order to perform high-throughput screening of the relationship between atomic structure and the resulting topological properties. We first discuss the electronic band structures from our DFTB model with and without SOC for the rhombohedral solid and two-dimensional bismuthene structure, with comparison to results from DFT. We then determine a many-body repulsive energy based on a minimal training set of DFT single-point calculations and molecular dynamics (MD) simulations, where we validate our model against the DFT-computed phonon density of states and energetic ordering of solid bismuth phases. Finally, we use DFTB to compute the bismuthene topological properties over a wide range of lattice constants and buckling parameters. Our results illustrate the ability of DFTB to help elucidate the structure-property relationships governing bismuth topological states that could help guide future materials development.

\section{Computational Methods}

Our DFT training data comprised electronic and physical properties from rhombohedral bismuth and bismuthene. All calculations were run with the Vienna ab initio Simulation Package (VASP)\cite{vasp,vasp2,vasp3}, using projector-augmented wave function (PAW) pseudopotentials\cite{Bloechl94,Kresse99} and the Perdew-Burke-Enzerhof (PBE) exchange-correlation functional.\cite{Perdew:1996} Our SOC and non-SOC calculations were run with identical inputs excluding the inclusion of spin-orbit coupling. We have used a planewave cutoff of 500~eV and an electron density convergence criterion of $10^{-6}$~eV for all calculations discussed here. For our band structure calculations, we have used a $20 \times 20 \times 20$ Monkhorst-Pack\cite{Monkhorst76, MP_erratum} k-point mesh for the two-atom rhombohedral primitive cell calculations, and a $15 \times 15 \times 1$ mesh for the two-atom buckled bismuthene primitive cell, unless otherwise noted. 

DFTB calculations were run with the DFTB$+$ code\cite{DFTB+_2025} using orbital-resolved SCC with a charge convergence criterion of $10^{-6}$. Band structure calculations were performed on identical primitive cells and k-point meshes to our DFT results. Thermal electron populations were determined using the Mermin functional\cite{Mermin65} with a Fermi-Dirac smearing factor of 0.05~eV. For this work, all DFTB SOC calculations have used the bismuth p-orbital constant of 1.645~eV, determined in Ref.~\citenum{Heine_DFTB_SOC}. 

Transferability of DFTB to multiple sets of systems and conditions can be optimized through confining potentials that are separately applied to the atomic orbitals and electron density. The DFTB Slater Koster Optimization (DSKO) code allows for these to be globally optimized to electronic band structure data.\cite{DSKO, cerium_dsko} DSKO computes the objective function through a decoding mask, defined by the energetic differences between bands at special k-points. In our work, we use separate Woods-Saxon confining potentials\cite{WoodsSaxon} for each angular momentum channel (5d, 6s, 6p) and the electron density. We have chosen to optimize the confining potential parameters to non-SOC DFT band structure calculations, only. This was performed in part as a test of the \emph{ab inito} transferability of the DFTB method, since the DFTB SOC constant is precomputed from high-level quantum chemical calculations and thus the wavefunction and density confinement can be determined separately. DSKO optimization was performed on the two-atom rhombohedral primitive cell, as well as the same configuration with the lattice vectors compressed by a factor of 0.98 in order to sample one higher density configuration. The initial rhombohedral primitive cell was taken from the Materials Project\cite{materials_project}. One bismuthene structure with an in-plane lattice constant $a=4.54$~\AA\ and a buckling distance $\Delta z=1.70$~\AA\ was also included to sample an out-of-equilibrium configuration. The confining potentials were determined using particle swarm optimization with 96 particles run over 50 iterations. Our final confining potential parameters are listed in the Supplementary Information.

\section{Results and Discussion}

\begin{table}[!htp]
\centering
\caption{DSKO decoding masks.}
\label{tab:dsko_decoding_masks}
\begin{tabular}{lcc}
\toprule
Structure & Special k-points & Decoding mask \\
\midrule
Rhombohedral Bi
    & $\Gamma, L, Z, B, X$
    & $[-2,-1,0,1]$ \\
Compressed rhombohedral Bi
    & $\Gamma, L, Z, B, X$
    & $[-2,-1,0,1]$ \\
Bismuthene
    & $\Gamma, M, K$
    & $[-2,-1,0,1]$ \\
\bottomrule
\end{tabular}
\end{table}

\begin{figure}[!htp]
    \centering
        \includegraphics[
            width=0.72\columnwidth
        ]{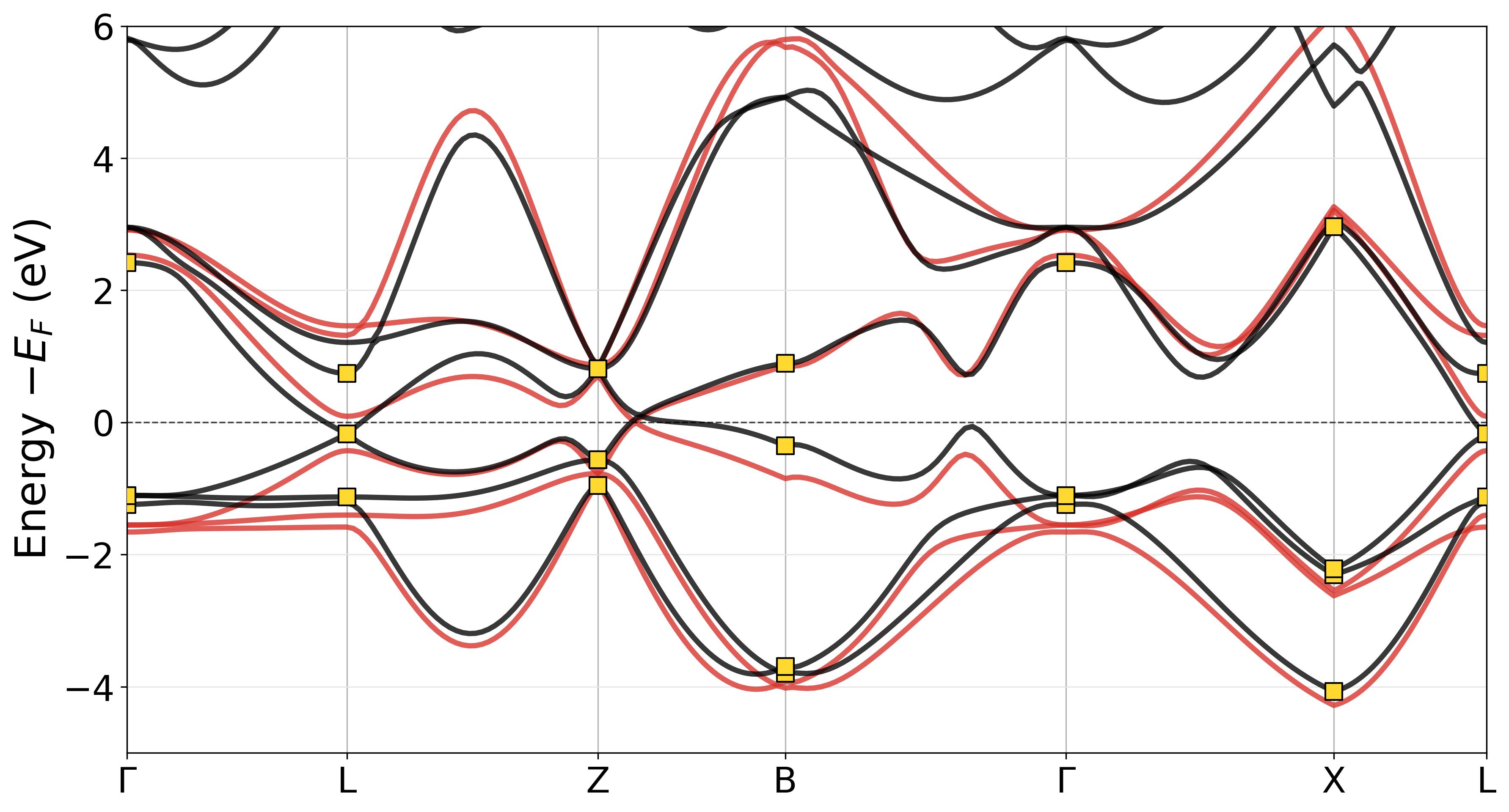}
    \caption{\label{fig:dsko_rhomb_nonscc} DSKO non-SOC optimization results for rhombohedral Bi with the decoding mask marked with yellow squares. Red lines correspond to DFTB and black to DFT.}
\end{figure}

We first evaluate our DSKO band structure optimization results for our non-SOC calculations (Table~\ref{tab:dsko_decoding_masks} and Figure~\ref{fig:dsko_rhomb_nonscc}). All special k-points sampled in the decoding mask are time-reversal-invariant momenta (TRIM) points, excluding the rhombohedral $B$-point and the bismuthene $K$-point. In general, our results for the uncompressed rhombohedral system show strong agreement with DFT, where the overall band structure features are well represented in our DFTB model. Orbital projections indicate that DFTB correctly identifies that the bands within a few eV of the Fermi energy have almost entirely p-orbital character (see Supplementary Information for more details). The band separation at the $\Gamma$-point from DFTB is slightly wider, though the degree of  band dispersion (electron delocalization) largely agrees with DFT when traversing the $\Gamma-L$ direction. We observe some discrepancy at the $L$-point, where DFT predicts a band crossing just below the Fermi energy whereas DFTB exhibits a small gap opening in the same region. In addition, DFTB predicts two bands at $\sim$1.5~eV close together on this scale that have larger separation and are closer to $\sim$~1~eV in DFT. However, DFTB yields a very similar band crossing and band gap closure along the $Z-B$ direction. The band separation at the $B$-point is slightly more broad in DFTB around the Fermi energy, and the excited states above 5~eV are more dispersive than those from DFT. However, the band structures largely agree at all other special k-points.  We observe similar agreement between DFTB and DFT for the compressed rhombohedral system (see Supplementary Information for details). 

\begin{figure}[!htp]
    \centering
    \subfloat{
        \includegraphics[
            width=0.72\columnwidth
        ]{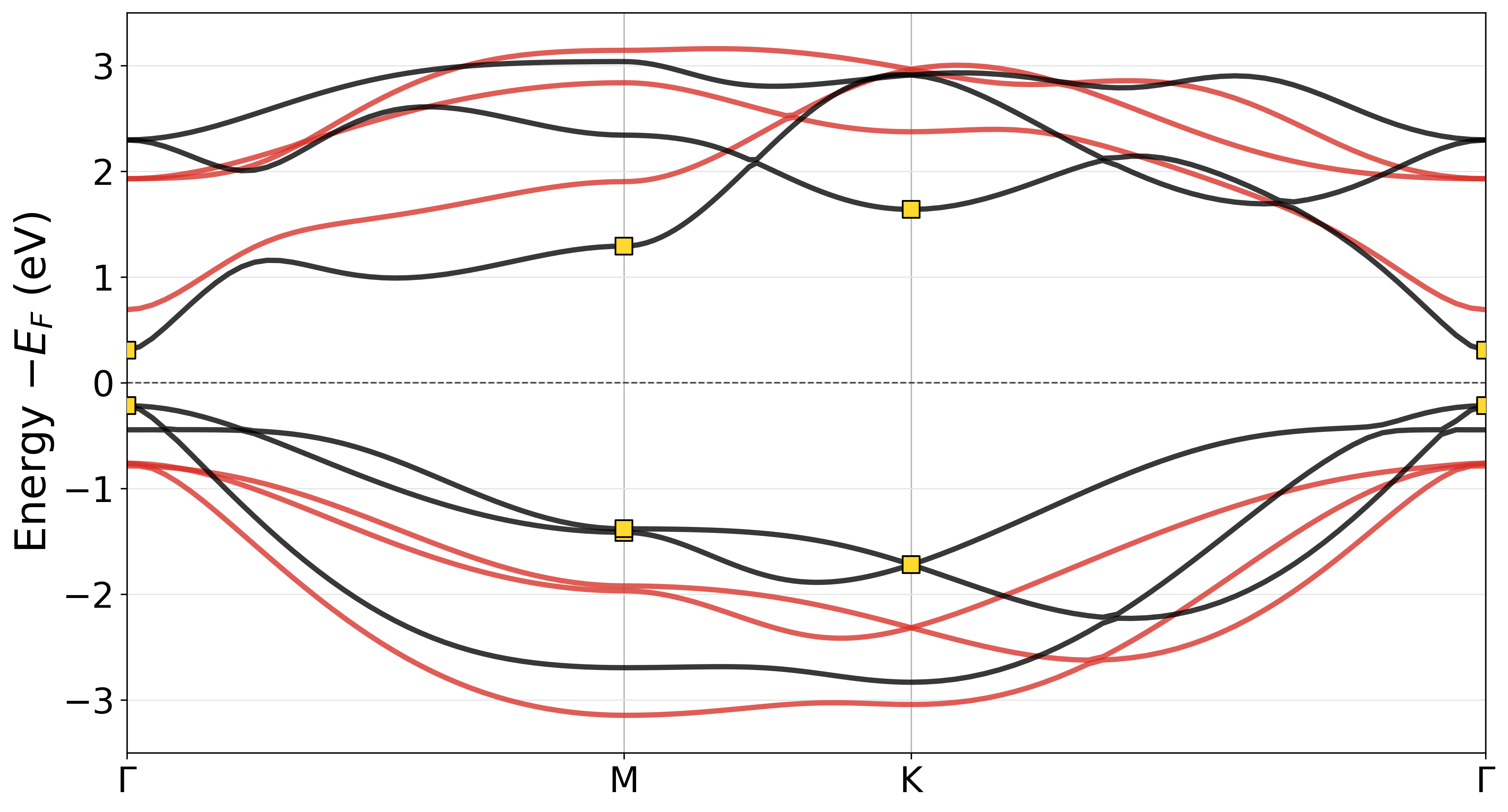}
    }
    \vspace{2mm}
    \subfloat{
        \includegraphics[
            width=0.75\columnwidth
        ]{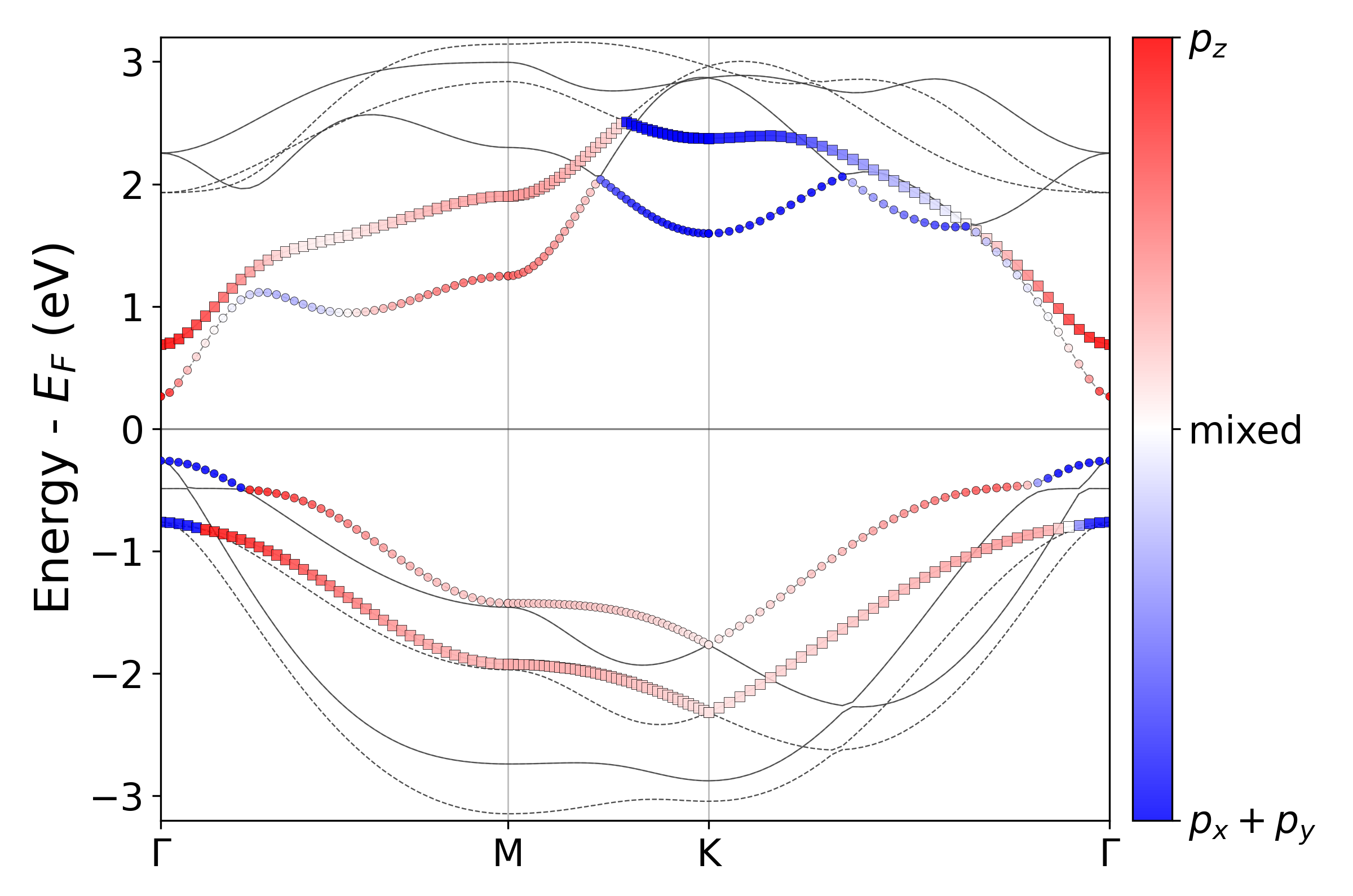}
    }
    \caption{DSKO non-SOC bismuthene results with marked decoding mask (top) and normalized p-orbital projections (bottom) for the VBM and CBM. Red lines in the top panel correspond to DFTB and black to DFT. Squares in the bottom panel correspond to the DFTB VBM/CBM and circles to DFT, with all other DFTB bands shown as dashed lines and DFT bands as solid. The color bar corresponds to the normalized $p_x + p_y$ (blue) versus $p_z$ (red) orbital character.}
    \label{fig:dsko_bis_nonscc}
\end{figure}

The non-SOC band structures from DFTB and DFT for bismuthene agree qualitatively, though with some differences (Figure~\ref{fig:dsko_bis_nonscc}). First, the band separation at the $\Gamma$-point about the Fermi energy is slightly wider in DFTB, though both methods predict a valence state that is predominantly $p_x + p_y$ and a conduction state that is largely $p_z$. However, the DFTB conduction band along the $\Gamma-M$ direction increases monotonically, whereas the DFT band is more oscillatory and exhibits a minimum and maximum along this path. This can be explained in part by the p-orbital projections, where the DFTB conduction band largely retains its $p_z$ character while the DFT band oscillates between $p_z$ and $p_x + p_y$ character, indicating a greater mixing of orbital character (dispersion) between those states. DFTB exhibits improved agreement with DFT past the mid-point of the band where both exhibit mixed $p_z/(p_x + p_y)$ character followed by predominantly $p_z$ character. Both methods show consistent band and p-orbital structure at the $M$ point, where $p_z$ character dominates in both the valence and conduction bands, with a larger band separation in the DFTB result. Finally, both DFTB and DFT show similar band structure at the $K$ point, where the valence band crossover shows mixed p-orbital character and the conduction band is distinctly $p_x + p_y$, though again the band separation in DFTB is somewhat larger than that from DFT. Overall, we find that DFTB reliably predicts the p-orbital character of the valence band and the bulk of the conduction band in the Brillouin zone path studied here, exhibiting similar regions of alternating $(p_x + p_y)/p_z$ character and band-crossing points along the $M-K$ and $K-\Gamma$ directions.  

\begin{figure}[!htp]
    \centering
        \includegraphics[
            width=0.72\columnwidth
        ]{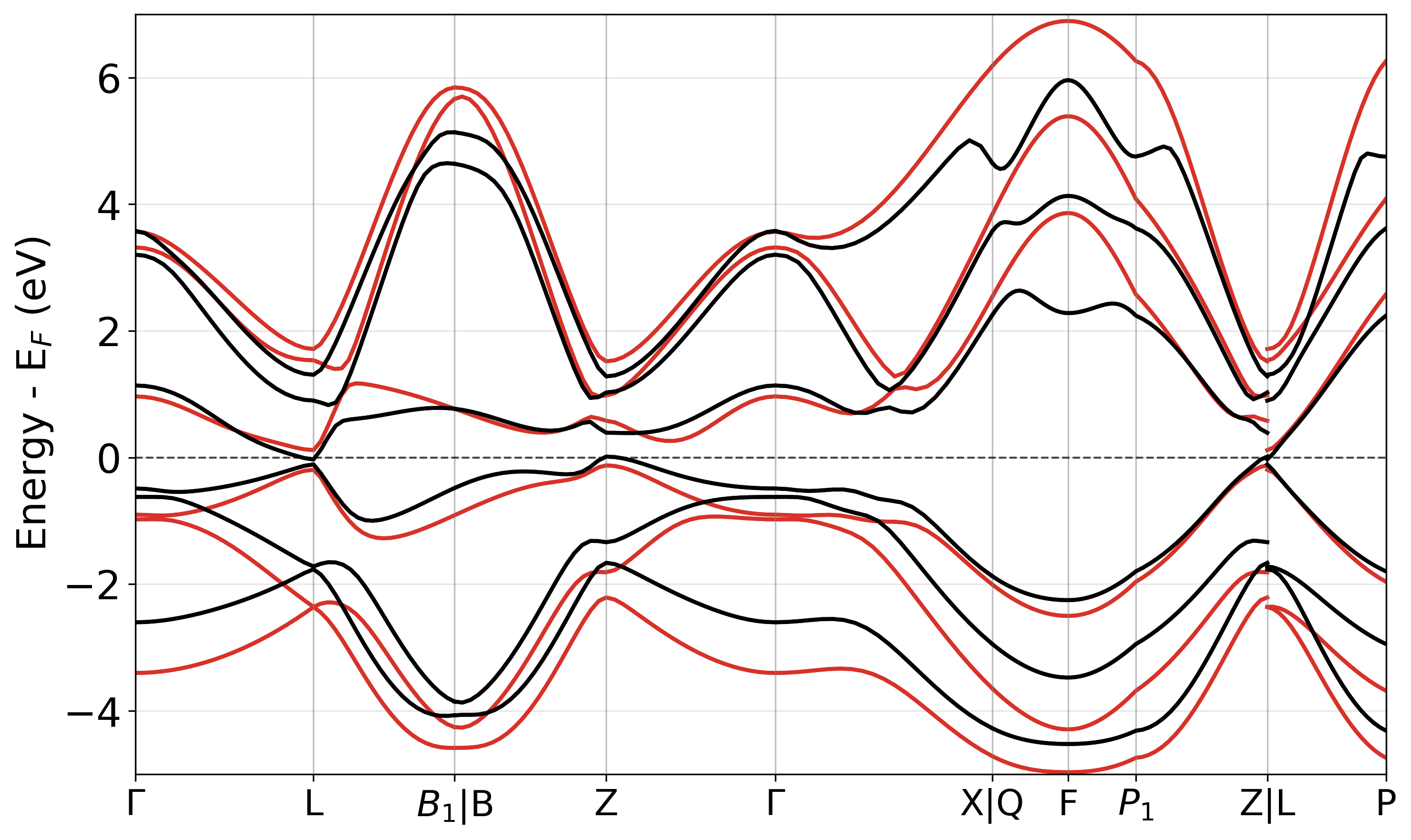}
    \caption{\label{fig:dsko_rhom_soc} Rhombohedral bismuth SOC results. Red lines correspond to DFTB and black to DFT. Higher energy conduction bands are removed for the sake of clarity.}
\end{figure}

We now examine the SOC band structure from DFTB for the rhombohedral bismuth ground state, where we observe overall improved agreement with DFT (Figure~\ref{fig:dsko_rhom_soc}). We observe similar bands at the $\Gamma$-point, though the DFTB band separation is slightly larger. However, we observe improved agreement at the $L$ point, where SOC splitting has caused a small gap to open at the Fermi energy in DFT which corresponds to the slightly larger gap observed in DFTB. DFTB yields low-lying valence and higher-energy conduction bands that are more dispersive than those from DFT along the $L-B_1$ and $B-Z$ directions. However, we note strong agreement between the two methods around the Fermi energy, where a band gap opens along $B-Z$ due to the presence of spin-orbit coupling. DFTB yields somewhat less dispersive conduction bands along the $\Gamma-X$, $Q-F$, and $F-P$ directions, where the DFT bands at $\sim$2~eV exhibit several minima and maxima that are not present in the DFTB result. Regardless, overall our results indicate that including SOC in our DFTB calculations improves prediction of the rhombohedral band structure, despite the fact that the confining potentials were optimized to the non-SOC DFT result. 

\begin{figure}[!htp]
    \centering
    \subfloat{
        \includegraphics[
            width=0.72\columnwidth
        ]{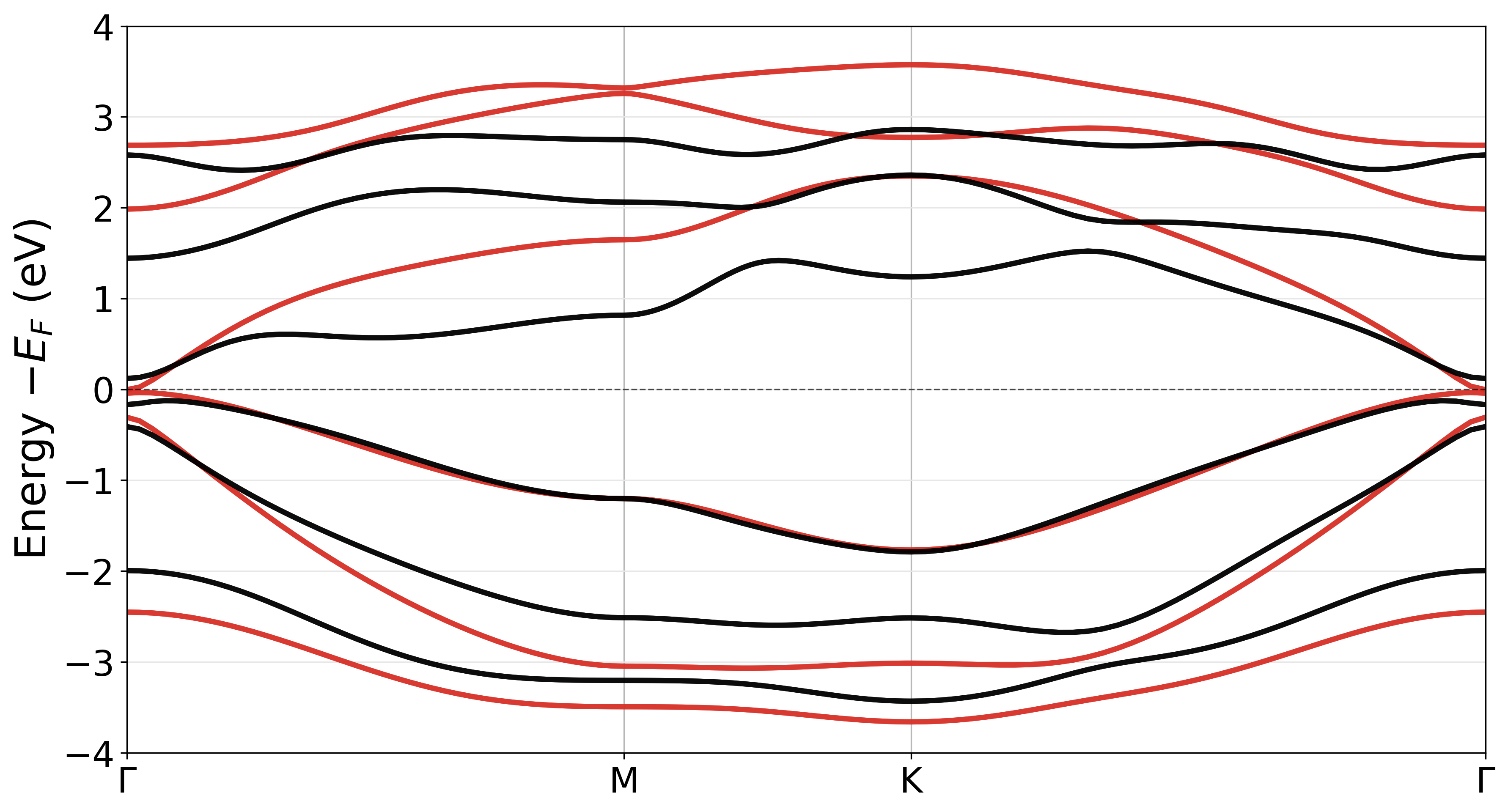}
    }
    \vspace{2mm}
    \subfloat{
        \includegraphics[
            width=0.72\columnwidth
        ]{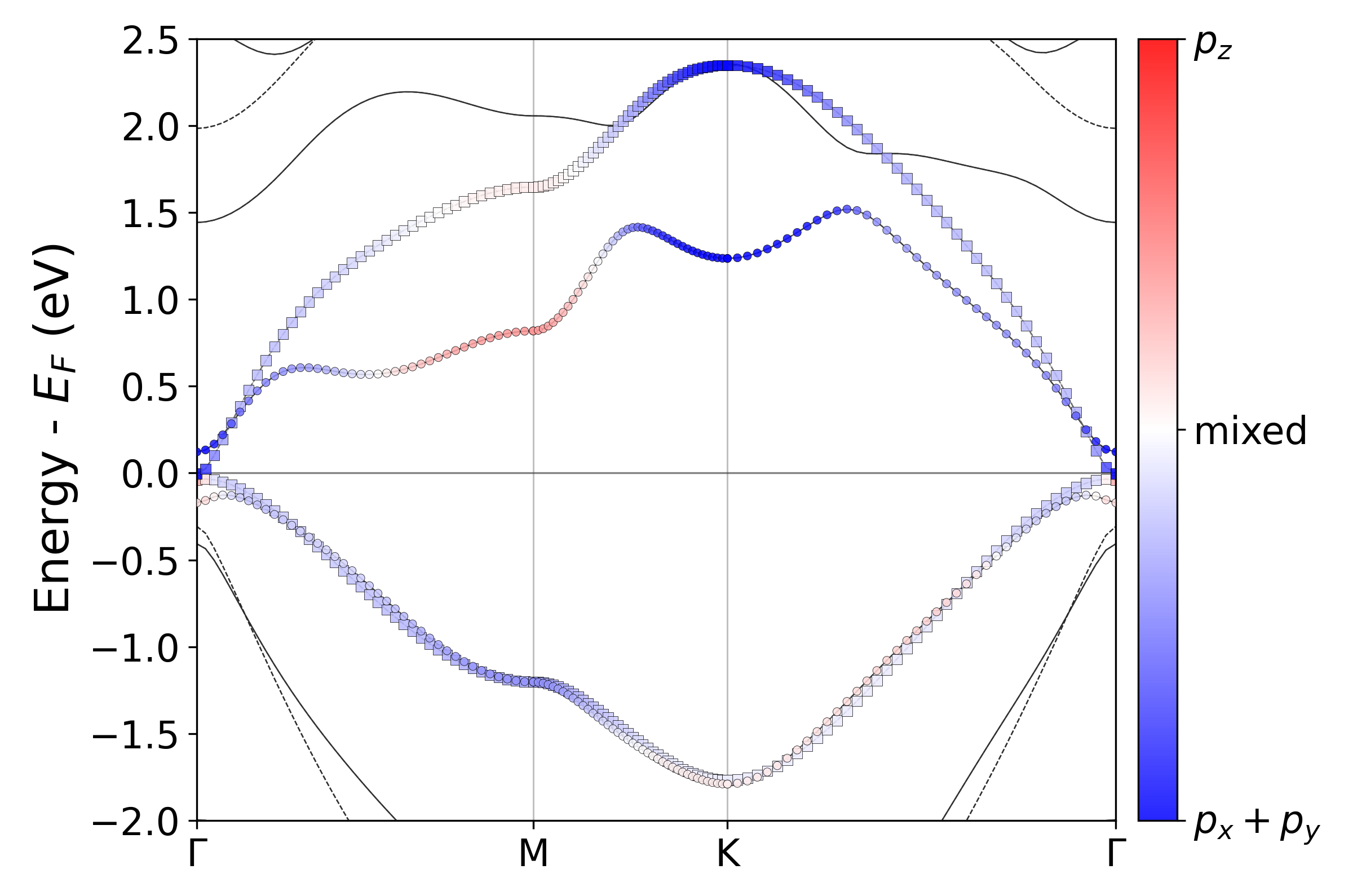}
    }
    \caption{Bismuthene SOC results (top) and normalized p-orbital projections for the VBM and CBM (bottom). Red lines in the top panel correspond to DFTB and black to DFT. Squares in the bottom panel correspond to the DFTB VBM/CBM and circles to DFT, with all other DFTB bands shown as dashed and DFT bands as solid. The color bar corresponds to the normalized $p_x + p_y$ (blue) versus $p_z$ (red) orbital character.}
    \label{fig:dsko_bis_soc}
\end{figure}

Examination of the bismuthene band structure computed with SOC indicates similarly improved agreement between DFTB and DFT. Again, the DFTB band separation is somewhat larger overall, though we note strong agreement with DFT at the $\Gamma$-point, where both methods exhibit a smaller band gap than the non-SOC result. In addition, we observe that DFTB yields a valence band maximum (VBM) that agrees closely with DFT throughout the entire sampled path. We also observe agreement for the p-orbital projections, with DFTB correctly predicting the p-orbital band inversion relative to the non-SOC result at $\Gamma$, where the VBM now has $p_z$ character and the conduction band minimum (CBM) has $p_x + p_y$. This orbital character is correctly reversed in DFTB at the $M$ TRIM point, where the VBM is now $p_x + p_y$ character and the CBM is $p_z$, though the CBM from DFTB has slightly less $p_z$ character than that from DFT. Band inversion has previously been used as a diagnostic tool for non-trivial topology in bismuthene in electronic structure calculations.\cite{Jin2015Bismuth, Pakdel2018_Sb_Bi_SOC, Liu2019Bismuthene, Li2021Bismuth} This gives preliminary indication that DFTB can correctly predict these properties and could serve as a tool for prediction of a range of topological structures.

We have created a ChIMES many-body repulsive energy in order to complete our model for structural predictions.\cite{Goldman_TiH2, Pham_DFTB_JPCL, Goldman_SEQM_2023, Dettori2025} Similar to previous work, we take an approach for $E_\mathrm{Rep}$ training that utilizes a minimal DFT training set.\cite{cerium_dsko} We have first run DFT-MD simulations on a rhombohedral supercell using the Materials Project-optimized structure (material ID mp-23152) as well as one configuration with the lattice vectors isotropically scaled to 0.98 and another with those scaled to 1.02 in order to sample a small range of non-equilibrium volumes. Here, our strategy is to use non-SOC DFT calculations for these simulations due to their increased computational efficiency, though all validation and production calculations will include SOC. Each simulation was run with a $3 \times 3 \times 3$ supercell (54 atoms) and $3 \times 3 \times 3$ k-point mesh for 7--8~ps, using a time step of 4.0~fs. We then sampled configurations every 400~fs in order to help minimize statistical correlations, yielding 59~geometries total from our MD simulations. We have included in our training set four additional configurations based on the same equilibrium supercell and k-point mesh where each bismuth atom has been displaced by a randomly oriented vector with a maximum magnitude of 0.1~\AA. This yields a total of 63 configurations or 10,458 data points, including atomic forces, diagonals of the stress tensor, and total energies, which is substantially smaller than previous machine-learning approaches for the repulsive energy.\cite{Margraf_GPR_2020, Pham_DFTB_JPCL, Yang_MLTB_2025, Tkatchenko_2026}

Our $E_\mathrm{Rep}$ training set was then determined by running non-SOC DFTB single-point calculations for each configuration without the repulsive energy, subtracting those values from the corresponding DFT result, and fitting ChIMES to the residual (see the Supplementary Information for the objective function and additional details regarding our optimizations). For our work, we have used a ChIMES potential with 12th order two-body interactions, eighth-order three-body interactions, and without four-body terms. We have used a weight of $5/n_\mathrm{atom}$ for the energies and a weight of 1.0 for everything else. The inner radial cutoff was set to 2.8~\AA\ (approximately the smallest interatomic distance sampled in our training set) and the outer cutoff to 4.0~\AA\ to include both nearest and next-nearest neighbor interactions. This resulted in RMS errors of 0.11~eV/\AA\ in the forces, 0.14~GPa in pressure, and 0.01~eV in the energy per atom.

\begin{figure}[!htp]
    \centering
        \includegraphics[
            width=0.72\columnwidth
        ]{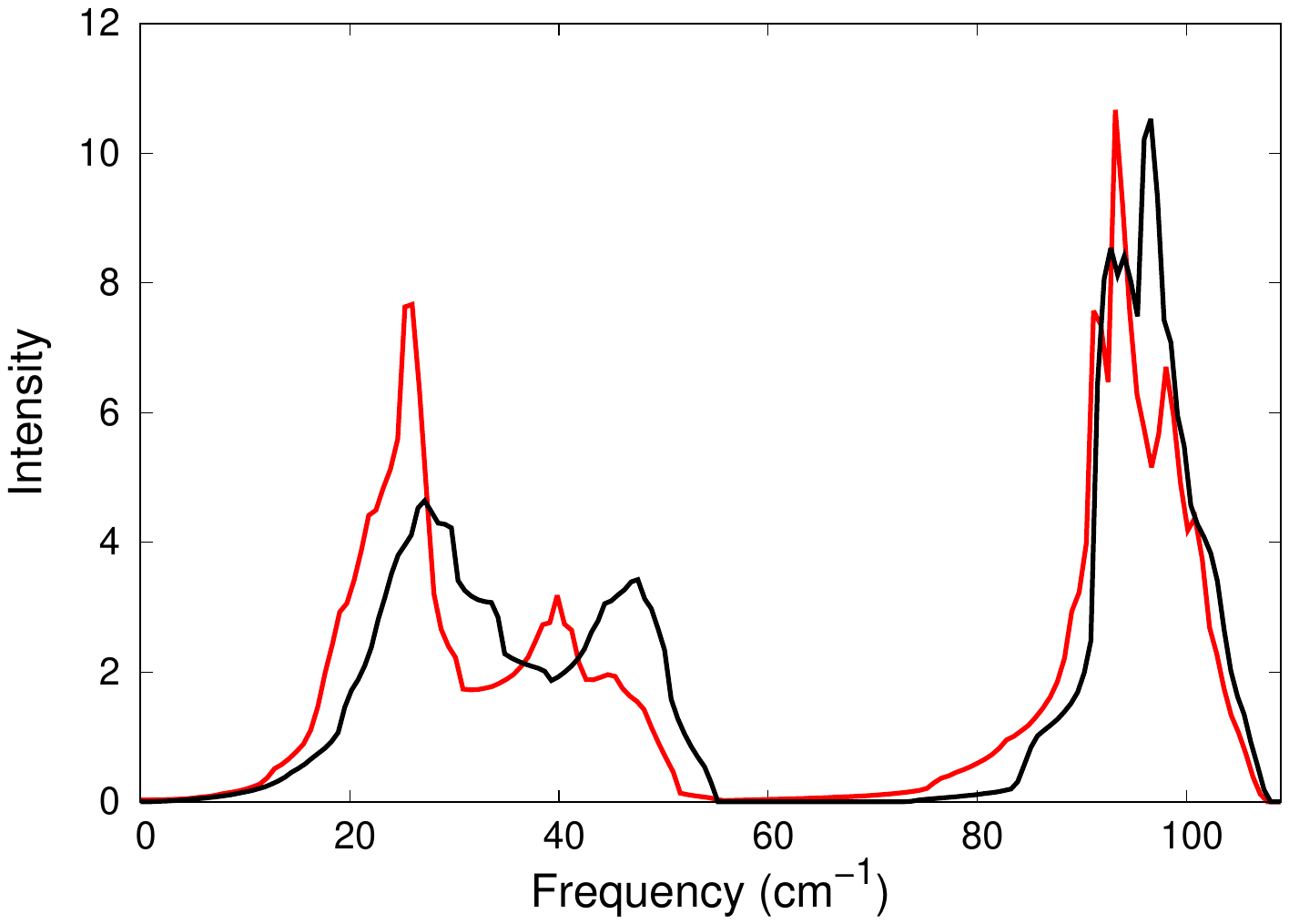}
    \caption{\label{fig:pdos} Spin-orbit coupling-computed total phonon DOS for rhombohedral bismuth. The red curve corresponds to  DFTB/ChIMES and the black to DFT.}
\end{figure}

We have validated our complete DFTB/ChIMES model through comparison to the total phonon density of states (phonon DOS) from DFT where we now have used SOC for both calculations. All of our results were determined using the phonopy code.\cite{phonopy-phono3py-JPCM, phonopy-phono3py-JPSJ} Overall, we observe close agreement between the two sets of results. DFTB/ChIMES reproduces the morphology of the doublet at frequencies below $\sim$50~cm$^{-1}$, where the peak at lower frequency has higher intensity than the second. However, the DFTB/ChIMES first peak is slightly higher in intensity and is more narrow than that from DFT. In addition, the DFTB/ChIMES doublet peaks are slightly redshifted to lower frequency relative to DFT, though these differences are relatively small (e.g., 5--10~cm$^{-1}$). Regardless, DFTB/ChIMES shows strong agreement with DFT for the manifold of peaks between $\sim$80-100~cm$^{-1}$, with only small differences in some peak intensities and an overall peak width that matches well. 

\begin{table}[!htp]
  \footnotesize
  \caption{Ground state energies relative to rhombohedral Bi ($\Delta E$) in eV/atom and nearest neighbor distances (NN) in \AA\ for the phases considered in this work.}
  \label{tab:bi_phase_comparison}
\begin{tabular}{c cc| cc| cc| cc}
\toprule
 & \multicolumn{2}{c}{rhombohedral} & \multicolumn{2}{c}{Bi-II} &  \multicolumn{2}{c}{simple cubic} &  \multicolumn{2}{c}{bcc} \\
\midrule
& $\Delta E$ & NN & $\Delta E$ & NN & $\Delta E$ & NN & $\Delta E$ & NN \\
DFTB/ChIMES     & 0.00 & 3.20 & 0.04 & 3.33 & 0.05 & 3.33 & 0.22 & 3.54  \\
DFT                      & 0.00 & 3.12 & 0.06 & 3.27 & 0.05 & 3.30 & 0.22 & 3.49 \\
\bottomrule
\end{tabular}
\end{table}

We further validate our DFTB/ChIMES model by computing the energies of three additional phases, not included in our training set, relative to the rhombohedral ground state: the body-centered cubic (bcc), simple cubic, and monoclinic Bi-II phases (Table~\ref{tab:bi_phase_comparison}). In this case, we have performed supercell optimizations using $3 \times 3 \times 3$ configurations of the conventional cell for the rhombohedral, Bi-II, and bcc phases using a $3 \times 3 \times 3$ k-point mesh. However, we have used a $5 \times 5 \times 5$ supercell with a $5 \times 5 \times 5$ k-point mesh for the simple cubic phase due to its lower density. We observe that our DFTB/ChIMES model is able to closely replicate the energetic ordering of these phases, where the largest discrepancy occurs with the Bi-II phase, which DFTB/ChIMES predicts to be 0.04~eV above the rhombohedral phase, instead of the result of 0.06~eV from DFT. We also determine consistent structural properties, where DFTB/ChIMES retains the symmetry of each phase with lattice constants that differ by no more than $\sim$3\% and angles by no more than $\sim$4\% (see Supplementary Information for details). Our DFTB/ChIMES model tends to predict slightly larger nearest neighbor (NN) distances overall, with the largest disagreement for the rhombohedral phase which deviates from the DFT result by $\sim$2.6\%. In addition, DFTB/ChIMES yields NN deviations from DFT for the Bi-II, simple cubic, and bcc phases of 1.8\%, 0.9\%, and 1.5\%, respectively. We thus observe a high degree of transferability for our approach, despite the fact that the $E_\mathrm{Rep}$ is only trained to data from a single phase. 

\begin{table*}[!htp]
\centering
\caption{Optimized bismuthene properties.}
\label{tab:bi_dftb_dft_comparison}
\begin{tabular}{c cc cc cc cc}
\toprule
& \multicolumn{2}{c}{$\Delta z$ (\AA)}
& \multicolumn{2}{c}{$d_{\mathrm{NN}}$ (\AA)}
& \multicolumn{2}{c}{$E_g$ (eV)}
& \multicolumn{2}{c}{$\mathbb{Z}_2$ invariant} \\
\cmidrule(lr){2-3}
\cmidrule(lr){4-5}
\cmidrule(lr){6-7}
\cmidrule(lr){8-9}
$a$ (\AA)
& DFTB/ChIMES & DFT
& DFTB/ChIMES & DFT
& DFTB/ChIMES & DFT
& DFTB/ChIMES & DFT \\
\midrule
4.30 & 1.80 & 1.78 & 3.07 & 3.05 & 0.17 & 0.54 & 1 & 1 \\
4.35 & 1.77 & 1.75 & 3.07 & 3.06 & 0.18 & 0.49 & 1 & 1 \\
4.40 & 1.74 & 1.73 & 3.08 & 3.07 & 0.19 & 0.43 & 1 & 1 \\
4.45 & 1.71 & 1.71 & 3.09 & 3.08 & 0.13 & 0.35 & 1 & 1 \\
4.50 & 1.68 & 1.68 & 3.09 & 3.10 & 0.04 & 0.26 & 1 & 1 \\
4.55 & 1.65 & 1.66 & 3.10 & 3.11 & 0.07 & 0.17 & 1 & 1 \\
4.60 & 1.62 & 1.64 & 3.11 & 3.12 & 0.12 & 0.07 & 1 & 1 \\
\bottomrule
\end{tabular}
\end{table*}

Next, we have used DFTB/ChIMES with SOC to predict the structural parameters and topological properties of buckled bismuthene. Here, we have held the in-plane lattice constant $a$ fixed over a range of 4.30 to 4.60~\AA\ and then optimized $\Delta z$, after which we have computed the nearest-neighbor value, global band gap, and the $\mathbb{Z}_2$ invariant. The $\mathbb{Z}_2$ invariant is determined by first computing the parity eigenvalues of the inversion operator matrix for the occupied subspace of the centrosymmetric structures at each TRIM point, namely $\Gamma$, $M_1$, $M_2$, and $M_3$. We then use the Fu-Kane parity criterion\cite{Fu_Kane_parity} by taking the product of one parity eigenvalue from each occupied Kramers pair at all four TRIM points, yielding values of $(-1)^\nu$ where $\nu$ can equal 0 or 1 and corresponds to $\mathbb{Z}_2$. A value of $\mathbb{Z}_2 = 0$ indicates trivial topology whereas $\mathbb{Z}_2 = 1$ indicates a topological insulating system (assuming the system has a global band gap). DFT values for the inversion-parity eigenvalues and $\mathbb{Z}_2$ invariants were computed using the Irrep software suite for validation.\cite{irrep_code} Both DFTB/ChIMES and DFT calculations were performed using a $21 \times 21 \times 1$ k-point mesh in this case. Additional details regarding DFTB calculation of the $\mathbb{Z}_2$ invariant are included in the Supplementary Information.

Our results indicate that DFTB/ChIMES largely reproduces the electronic and structural properties of optimized buckled bismuthene for all geometries studied here (Table~\ref{tab:bi_dftb_dft_comparison}). We observe that DFTB/ChIMES predicts the correct inverse correlation between the $a$ lattice parameter and $\Delta z$, with $\Delta z$ decreasing from 1.80~\AA\ to 1.62~\AA\ as $a$ increases from 4.30 to 4.60~\AA, compared to DFT computed $\Delta z$ values of 1.78 and 1.64~\AA\ over the same $a$ range (i.e., agreement within approximately 1\%). This translates to similar agreement for the Bi nearest neighbor distance for each structure, where the DFTB/ChIMES NN values increase from 3.07 to 3.11 \AA\, while DFT yields an increase from 3.05 to 3.12 \AA. There is some disagreement for the electronic band gap, where DFTB/ChIMES has a tendency to underestimate the gap for many of these structures. For example, DFTB/ChIMES yields a band gap of 0.17~eV for the $a=4.30$~\AA\ structure, compared to the value of 0.54~eV from DFT. This can be explained in part by the fact that the VBM and CBM p-orbitals from DFTB/ChIMES tend to be less dispersive than those from DFT (i.e., lower band curvature). This can result in more localized orbitals and less hybridization between p-orbital types, which in turn will cause the DFTB-computed band gap to be lower overall. However, DFTB/ChIMES yields the correct trend, where the gap tends to decrease with increasing value of $a$, and we observe band gaps of 0.07 and 0.12~eV at $a$ values of 4.55 and 4.60~\AA, compared to the DFT values of 0.17 and 0.07~eV. 

Regardless, DFTB/ChIMES correctly predicts that each optimized geometry is a topologically non-trivial insulator, with $\mathbb{Z}_2$ values of 1 throughout. Each structure can be computed with DFTB/ChIMES in a fraction of the time required by DFT. For example, a single optimization step with DFTB for the $a=4.60$~\AA\ geometry required 1.2 seconds on a single Intel Xeon Ice Lake Gold 2.8 GHz processor (or 1.2 CPU-seconds), compared to approximately 110 seconds on 24 processors for DFT (2640 CPU-seconds). Our DFTB/ChIMES model can thus deliver $\sim2200\times$ improvement in computation efficiency, allowing for a detailed exploration of the bismuthene topological phase space with relatively small computational resources. 

We now use our DFTB/ChIMES model to compute the electronic properties of bismuthene over a wide range of $a$ and $\Delta z$ structural parameters, including the $\mathbb{Z}_2$ invariant and global band gap. We also evaluate a projection-based orbital-inversion descriptor (validated flip score), computed at the $\Gamma$-point 

\begin{equation}
F_\mathrm{validate} = F_\mathrm{SOC} F_\mathrm{nonSOC}.
\end{equation}

\noindent Here, we define the following band population parameters

\begin{equation}
F_\mathrm{SOC} = \frac{f^\mathrm{SOC,HO}_{p_z} + f^\mathrm{SOC,LU}_{p_x + p_y}}{2}, F_\mathrm{nonSOC} = \frac{f^\mathrm{nonSOC,HO}_{p_x + p_y} + f^\mathrm{nonSOC,LU}_{p_z}}{2}.
\end{equation}

\noindent In this case, $f^\mathrm{SOC,HO}_{p_z}$ corresponds to the $p_z$ electronic population of the highest occupied band (HO) computed with SOC normalized to that band's total p-orbital weight, and $f^\mathrm{SOC,LU}_{p_x + p_y}$ to the $p_x + p_y$ normalized population of the lowest unoccupied (LU) band. Similarly, $f^\mathrm{nonSOC,HO}_{p_x + p_y}$ and $f^\mathrm{nonSOC,LU}_{p_z}$ correspond to the non-SOC normalized $p_x + p_y$ population of the HO and normalized $p_z$ population of the LU. The validated flip score thus measures the band inversion induced by SOC weighted by how close the non-SOC bands are to the expected uninverted populations. A perfect score of 1 is achieved only if the SOC result yields p-orbital projections exclusively with $p_z$ character for the HO and $p_x + p_y$ character for the LU, and vice-versa for the non-SOC result. 

\begin{figure}[!htp]
    \centering
    \subfloat[$\mathbb{Z}_2$ invariant. \label{fig:z2-contour}]{
        \includegraphics[
            width=0.45\columnwidth
        ]{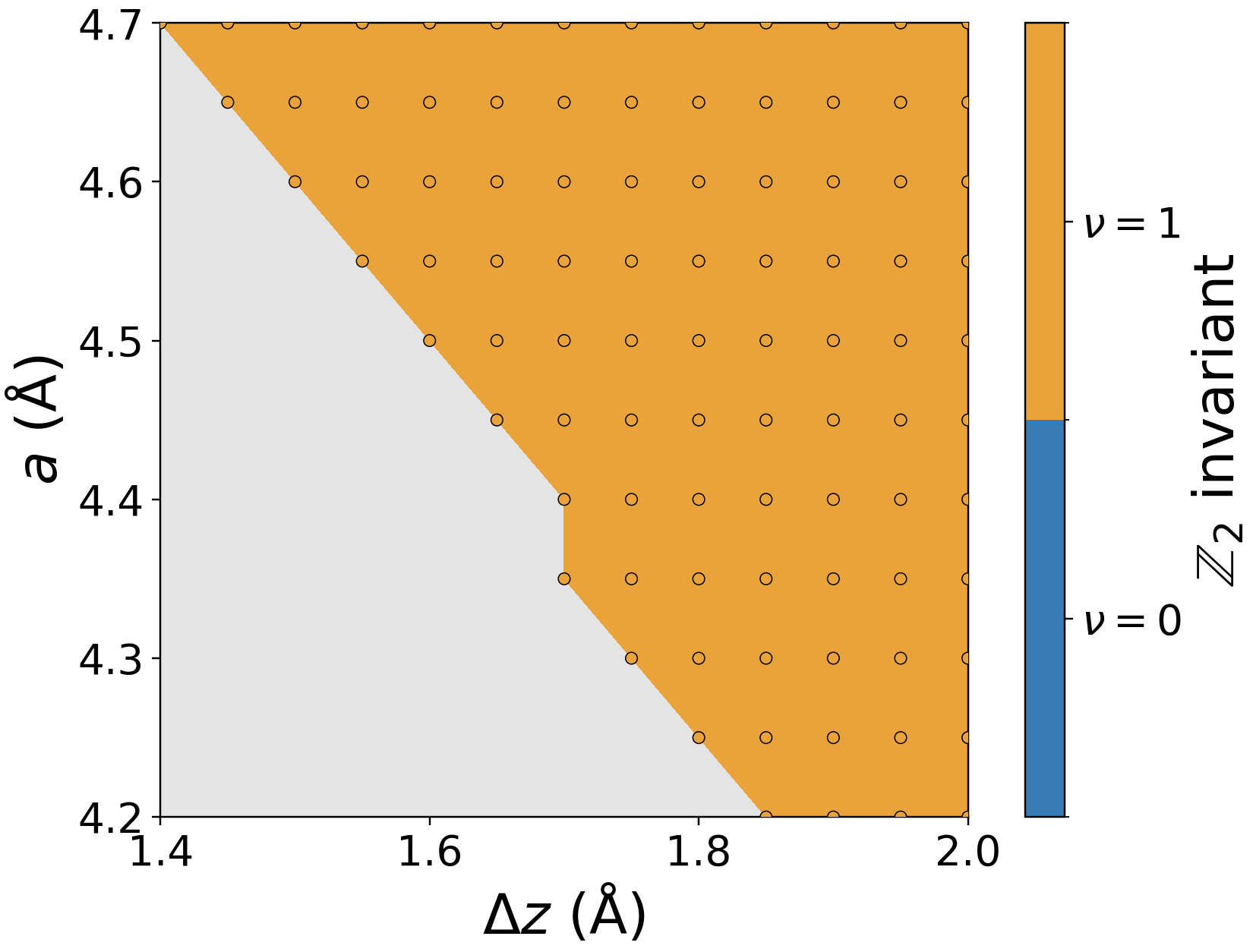}
    }
    \vspace{2mm}
    \subfloat[Validated flip score.\label{fig:flip-contour}]{
        \includegraphics[
            width=0.45\columnwidth
        ]{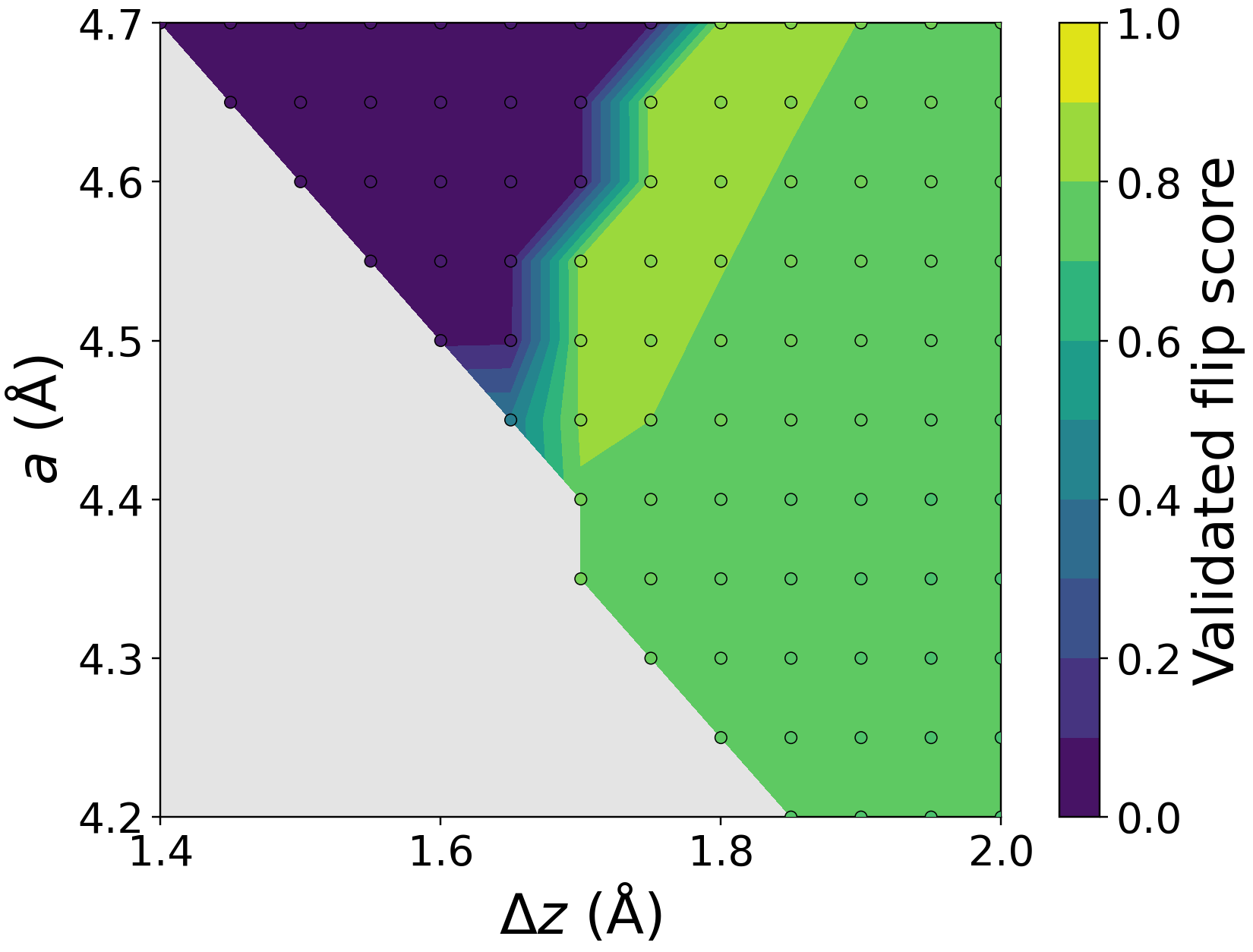}
    }
    \vspace{2mm}
    \subfloat[Band gap.\label{fig:gap-contour}]{
        \includegraphics[
            width=0.45\columnwidth
        ]{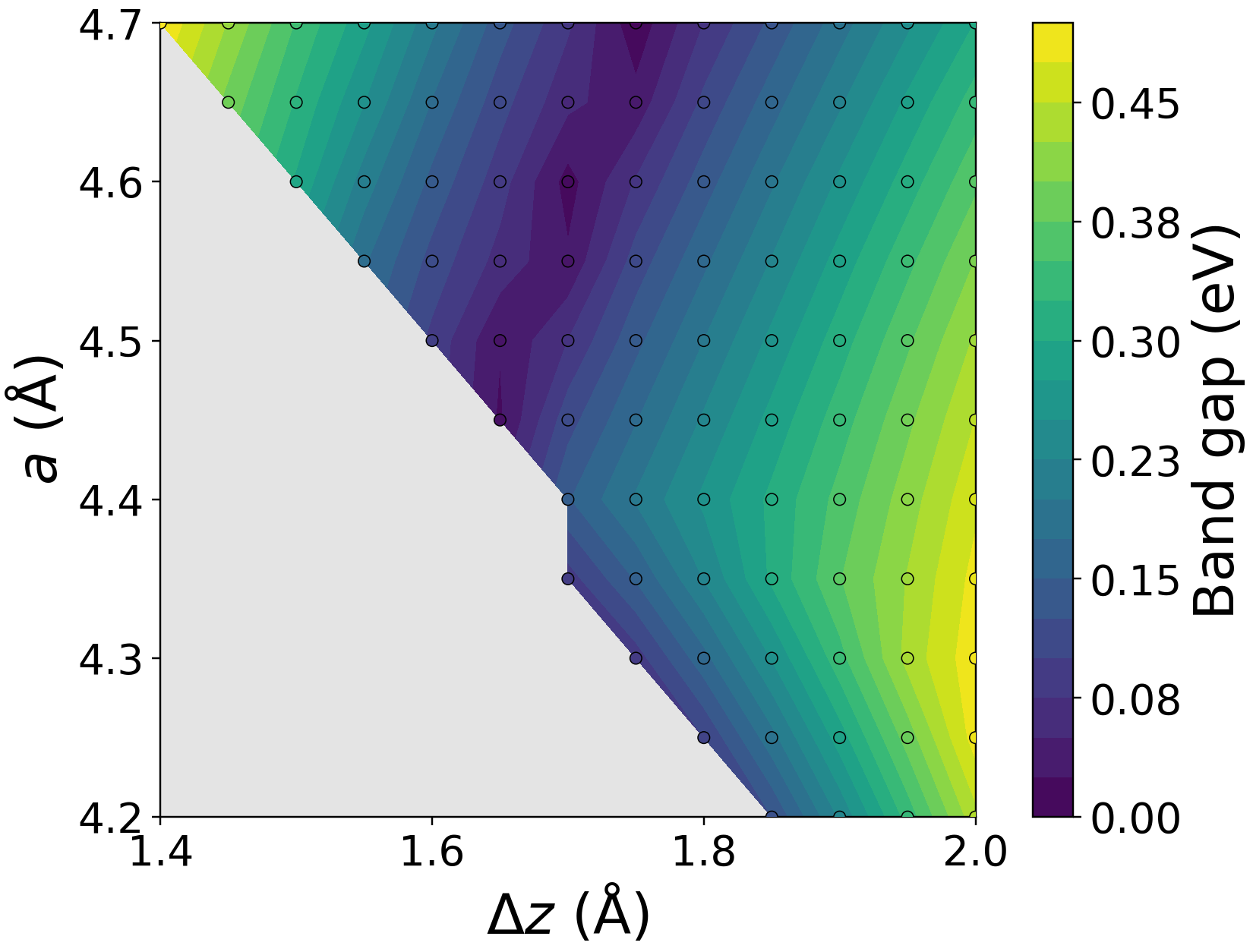}
    }
    \caption{Contour plots of the (a) $\mathbb{Z}_2$ invariant, (b) validated flip score, and (c) global electronic band gap as a function of bismuthene structural parameter sweep. Excluded points correspond to Bi-Bi distances below that probed in our DSKO optimization training data.}
    \label{fig:contours}
\end{figure}

Our results for all three metrics yield contour maps indicating how the electronic properties of bismuthene change as a function of structural parameters (Figure~\ref{fig:contours}). We have excluded from our analysis bismuthene structures that probe Bi interatomic distances smaller than those sampled in our DSKO confining potential optimizations (i.e., 3.03~\AA), as our DFTB/ChIMES calculations are unvalidated for those points. We observe that all of the structures generated in our parameter sweep are topologically non-trivial, with $\mathbb{Z}_2$ classification values of 1. In contrast, our validated flip score shows a region of the sampled structural space in which the p-orbital character of the bands remains uninverted, at values of $a$ above $\sim$4.45~\AA\ but below values of $\Delta z$ of 1.75~\AA. The individual Fu-Kane parity products for each TRIM point remain unchanged throughout the entire bismuthene structural scan (e.g., values of $-1$ for the $\Gamma$-point occupied manifold and values of 1 for all other TRIM points), even where the validated flip score is equal to zero. This dividing surface bears close correspondence to the region of band gap narrowing, extending from approximately ($a$=4.45~\AA, $\Delta z = 1.65$~\AA) to ($a$=4.70~\AA, $\Delta z = 1.75$~\AA). The $\Gamma$-point frontier band p-orbital character is consequently insufficient on its own as an order parameter for non-trivial topology, where we observe that its value can change significantly without affecting the $\mathbb{Z}_2$ classification or TRIM parity product. These results also indicate the possibility of using structural deformations to tune both the frontier orbital character and system band gap of bismuthene without losing its non-trivial topology, yielding the possibility of topology-preserving electronic band gap engineering.

\begin{table}[t]
\centering
\caption{Comparison of properties for selected fixed bismuthene geometries.}
\label{tab:bi_fixed_geometry_comparison}
\begin{tabular}{cc cc cc}
\hline
& & \multicolumn{2}{c}{Band gap (eV)}
& \multicolumn{2}{c}{Validated flip score at $\Gamma$} \\
\cmidrule(lr){3-4}
\cmidrule(lr){5-6}
$a$ (\AA) & $\Delta z$ (\AA)
& DFTB/ChIMES & DFT
& DFTB/ChIMES & DFT \\
\midrule
4.60 & 1.50 & 0.30 & 0.20 & 0.06 & 0.07 \\
4.55 & 1.70 & 0.03 & 0.24 & 0.83 & 0.79 \\
\bottomrule
\end{tabular}
\end{table}

We perform a final validation of our DFTB/ChIMES model against DFT by comparing results for the band gap and validated flip score for bismuthene structures at identical ($a$, $\Delta z$) points (Table~\ref{tab:bi_fixed_geometry_comparison}). We observe strong agreement between the two methods at (4.60, 1.50), where DFTB/ChIMES yields a band gap of 0.30~eV and validated flip score of 0.06 (indicating an HO state at the $\Gamma$-point predominantly with $p_x + p_y$ character), compared to band gap of 0.20~eV and a score of 0.07 from DFT. We observe similarly strong agreement at (4.55, 1.70) for the validated flip score, where DFTB yields a value of 0.83 and for DFT a value of 0.79 (the HO state that is predominantly $p_z$ in character). However, DFTB/ChIMES underestimates the band gap with a value of 0.03~eV, compared to the DFT value of 0.24~eV. This could be due to DFT exhibiting the band gap narrowing at a slightly different set of bismuthene parameters. Regardless, DFTB/ChIMES shows strong validation for bismuth-related structural, electronic, and topological properties over a broad range of configurations.

\section{Conclusions}

DFTB/ChIMES has proven to be a robust approach for predicting SOC-induced topological properties of bismuth with DFT-like accuracy while yielding over a factor of $10^3$ increased computational efficiency. Our model was optimized only to the electronic band structure for the rhombohedral and bismuthene phases and 63 configurations of rhombohedral phase MD and single point DFT data. It is able to accurately predict the rhombohedral phonon density of states and energetic ordering of several additional bismuth phases. In particular, we use it to perform high-throughput calculations with DFTB/ChIMES to predict structure-dependent maps of topological classification, the band gap, and orbital character of bismuthene as a function of the lattice constant and buckling parameter. Our results show that bismuthene has a region of near band gap closure that is saddled by topological insulating structures with the expected $(p_x + p_y)/p_z$ band inversion for larger structural parameters and those with non-trivial topology but with a lack of band inversion at lower values. This is indicative of possible structures where the Fu-Kane parity criterion does not change for the TRIM points, even though the p-orbital projections experience significant modification. 

Our efforts emphasize how bismuthene could be the source of a topology-preserving material with a tunable band gap or electronic transport properties. This could open the door for future studies where the bulk insulating gap could be modified without sacrificing the underlying topological phase. In addition, applied strain could be used to create topological insulators with varying electronic properties where the gap and valence/conduction band orbital character can be modified independently. These approaches could aid in the development of next generation bismuth-containing materials, where the optical, spin, and electronic transport properties could be refined by changing substrates or interfaces without changing the underlying topological states.  

\begin{acknowledgments}

This was work performed under the auspices of the U.S. Department of Energy
by Lawrence Livermore National Laboratory under Contract DE-AC52-07NA27344.

\end{acknowledgments}

\section*{Data Availability}

DFTB$+$ skf files and ChIMES repulsive energies for bismuth are available for download at https://github.com/nirgoldman/dftb\_bismuth.git.
\\
\\
\emph{Supporting Information:} optimized DSKO output for the DFTB$+$ Slater-Koster interaction files, the ChIMES repulsive energy optimization details, rhombohedral band structure p-orbital projections, compressed rhombohedral phase band structures, lattice optimization results from ChIMES/DFTB for different bismuth solid phases, details of the DFTB $\mathbb{Z}_2$ invariant calculations.

\newpage

\bibliography{library}

\end{document}


\newpage

\section{ChIMES Methodology}
An in-depth review of ChIMES can be found in Ref.~\citenum{Goldman_SEQM_2023}. Briefly, in the ChIMES formalism, we represent each of the terms in our n-body energy expansion as a linear combination of Chebyshev polynomials. Chebyshev polynomials of the first kind of order $m$ are defined by the expression $T_m \left(\mathrm{cos} \, \theta \right) = \mathrm{cos}\left(m \theta\right)$, more commonly written as $T_m(x)$, where $x = \mathrm{cos}\, \theta$ and thus exists over the range $\left[-1,1\right]$. The derivatives of Chebyshev polynomials of the first kind are related to Chebyshev polynomials of the second kind $U_m(x)$ by the expression $\mathrm{d}T_m/\mathrm{d}x = m U_{m-1}$, where $U_m\left( \mathrm{cos} \, \theta \right) = \mathrm{sin} \left[ \left(m+1\right) \theta \right]/\mathrm{sin} \, \theta$.

We write the two-body (2B) energy term as the following expression:

\begin{equation}
        {}^2\!E_{i_1 i_2} = f_{\mathrm{p}}\left(r_{i_1 i_2}\right) + f^{e_{i_1} e_{i_2}}_{\mathrm{c}}\left(r_{i_1 i_2}\right)
             \sum^{\mathcal{O}_{2}}_{m=1} C_m^{e_{i_1} e_{i_2}} T_m (s^{e_{i_1} e_{i_2}}_{i_1 i_2})
        \label{eqn:FF2B}
\end{equation}

\noindent $C_{m}^{e_{i_1} e_{i_2}}$ is an optimized coefficient for the interaction between atom types $e_{i_1}$ and $e_{i_2}$, taken from the set of all possible element types, $\{\boldsymbol{e}\}$. All $C_{m}^{e_{i_1} e_{i_2}}$  are permutationally invariant. $T_{m}\left(s^{e_{i_1} e_{i_2}}_{i_1 i_2}\right)$ represents a Chebyshev polynomial of order $m$, and $s^{e_{i_1} e_{i_2}}_{i_1 i_2}$ is the pair distance transformed to occur over the interval $[-1,1]$ using a Morse-like function\cite{WHBB-2009,WHBB-2011}. For that coordinate transform, $s^{e_{i_1} e_{i_2}}_{i_1 i_2} \propto \mathrm{exp}\left(-r_{i_1 i_2}/\lambda_{e_1 e_2}\right)$ and $\lambda_{e_1 e_2}$ is an element-pair distance scaling constant, frequently set to the first peak in a radial distribution function. Further details are discussed in Ref.~\citenum{Lindsey17}. The term $f^{e_{i_1} e_{i_2}}_\mathrm{c}(r_{i_1 i_2})$ is a Tersoff cutoff function\cite{Tersoff88} which smoothly varies to zero up to a predefined maximum cutoff distance. In order to prevent sampling of $r_{i_1 i_2}$ distances below those sampled in our training set, we introduce a smooth repulsive penalty function $f_\mathrm{p} (r_{i_1 i_2})$ that is non-zero for distances close to the inner cutoff of the Chebyshev polynomials.

Higher-body (e.g., greater than two-body) orthogonal polynomials can be created by defining a cluster of size $n$ and taking the product of the Chebyshev polynomials derived from the constituent ${n \choose 2}$ unique pairs. For example, the three-body polynomials will be products of ${3 \choose 2} = 3$ two-body polynomials. We thus write the ChIMES three-body (3B) energy as the following:

\begin{multline} 
{}^3 E_{i_1 i_2 i_3} = f^{e_{i_1} e_{i_2}}_{\mathrm{c}}\left(r_{i_1 i_2}\right) f^{e_{i_1} e_{i_3}}_{\mathrm{c}}\left(r_{i_1 i_3}\right) f^{e_{i_2} e_{i_3}}_{\mathrm{c}}\left(r_{i_2 i_3}\right) \\ 
 \sum^{\mathcal{O}_{3}}_{m=0} \sum^{\mathcal{O}_{3}}_{p=0} {\sum^{\mathcal{O}_{3}}_{q=0}}^\prime 
 C^{e_{i_1} e_{i_2} e_{i_3}}_{mpq} T_{m}\left(s^{e_{i_1} e_{i_2}}_{i_1 i_2}\right) T_{p}\left(s^{e_{i_1} e_{i_3}}_{i_1 i_3}\right) T_{q}\left(s^{e_{i_2} e_{i_3}}_{i_2 i_3}\right).
        \label{eqn:FF3B}
\end{multline}

\noindent We thus compute a triple sum for the product of the $i_1 i_2$, $i_1 i_3$, and $i_2 i_3$ pairwise polynomials. These are computed up to a predefined order ($\mathcal{O}_{3}$) for each three-body polynomial and then multiplied by a single coefficient, $C^{e_{i_1} e_{i_2} e_{i_3}}_{mpq}$, that is permutationally invariant for each set of polynomial orders and atom types. The primed sum in Equation~\eqref{eqn:FF3B} indicates that only terms for which two or more of the $m,p,q$ polynomial powers are greater than zero are included in order to guarantee that all atoms in the cluster interact through Chebyshev polynomials.  
The expression for ${}^3\!E_{i_1 i_2 i_3}$ also contains the $f_{\mathrm{c}}$ smoothly varying cutoff functions for each constituent pair distance. The penalty function $f_p$ is not included here.

Higher-body terms are included in ChIMES in a similar fashion. For example, four-body (4B) terms are often included in ChIMES optimizations\cite{Pham_HN3_2021}, where ${}^4\!E_{i_1 i_2 i_3 i_4}$ is now determined from the sum over the product of the ${4 \choose 2} = 6$ constituent pairwise polynomials multiplied by a single permutationally invariant coefficient. In the determination of permutational invariance for an arbitrary number of bodies, it is important to realize that the atom indices and atom types are permuted, which then implies a corresponding permutation of the bond distance indices in $C$.   

Similar to other machine learning atomic interaction potentials, ChIMES uses the method of force matching\cite{Ercolessi94} to determine the interaction parameters.
In force matching, a training set of quantum simulations is generated with differing configurations.  For each configuration in the training set, the quantum mechanical
energy, atomic force, and stress (for condensed phase configurations) are calculated. In addition, weights can be used when matching to forces, energies, and stresses, due to the differing physical units and number of parameters per configuration.  In this work, we use a weight of $5/n_{a}$ for the energies ($n_a$ is equal to the number of atoms in a given configuration), and a value of one for the forces. Once weights are determined, an objective function for optimization may be defined as follows:

\begin{footnotesize}
\begin{equation}
        F_{\mathrm{obj}}  =  \frac{1}{N_d} \sum_{\tau=1}^M  \left(\sum_{i=1}^{N_\tau} \sum_{\alpha=1}^{3} \left(w_{\mathrm{F}}\Delta \mathrm{F}_{\tau_{\alpha_i}} \right)^{2} + \sum_{\alpha=1}^3 \sum_{\beta \le \alpha}\left(w_{\sigma}\Delta\sigma_{\tau_{\alpha\beta}}\right)^{2} + \left(w_{\mathrm{E}} \Delta E_{\tau}\right)^{2}
\right)
.
\label{eqn:rmse}
\end{equation}
\end{footnotesize}

\noindent Here, $\tau$ corresponds to a specific training set configuration, $i$ is the atomic index, and $\alpha$ and $\beta$ are the Cartesian directions. 
$M$ is the total number of configurations in the training set and $N_d = 3\sum_\tau N_\tau + 4 M$ is the total number of data entries (three stress tensor components and one energy value per configuration).  In addition, $\Delta F_{\tau_{\alpha_i}} = F^{\mathrm{ChIMES}}_{\tau_{\alpha_i}} - F^{\mathrm{DFT}}_{\tau_{\alpha_i}}$, $\Delta\sigma_{\tau_{\alpha\beta}} = \sigma^{\mathrm{ChIMES}}_{\tau_{\alpha\beta}} - \sigma^{\mathrm{DFT}}_{\tau_{\alpha\beta}}$, 
and $\Delta E_{\tau} = E^{\mathrm{ChIMES}}_{\tau} - E^{\mathrm{DFT}}_{\tau}$.  The value $w_{\mathrm{F}}$ is the weight for forces, $w_{\sigma}$ for stresses, and $w_{\mathrm{E}}$ for energies. 

Optimal ChIMES parameters are determined by solving 
\begin{equation}
\frac{\partial F_{\mathrm{obj}}}{\partial C_I} = 0,
\label{eqn:optF}
\end{equation}
for all $I$, where $I$ is a combined index of the permutationally unique coefficient. 
Optimizing $F$ is equivalent to solving the overdetermined matrix equation 
\begin{equation}
\boldsymbol{w} \boldsymbol{A} \boldsymbol{C} = \boldsymbol{w} \boldsymbol{B}.
\label{eqn:linear}
\end{equation}
The matrix $\boldsymbol{A}$ corresponds to the derivatives of the ChIMES energy, stress, or force expression with respect to the fitting coefficients.  The column vectors $\boldsymbol{C}$ and $\boldsymbol{B}$ correspond to the ChIMES coefficients to be optimized and the numerical values for the training data, respectively. The diagonal matrix $\boldsymbol{w}$ is comprised of the weights to be applied to the
elements of $\boldsymbol{B}$ and rows of $\boldsymbol{A}$. Solution to this linear least-squares problem can be performed using a number of different optimization algorithms, discussed in Ref.~\citenum{Goldman_SEQM_2023}.

\section{ChIMES parameters and fitting results}
For the ChIMES repulsive energy discussed in this work, we have used a minimum radial cutoff of 2.80 \AA\, which was the minimum bismuth nearest neighbor distance sampled in our training set.  We have set the ChIMES maximum cutoff to a value of 4.00 \AA\, which ensures inclusion of second-nearest neighbor interactions for the molecular dynamics simulations discussed in the manuscript. Optimizations were performed with the LASSO/LARS algorithm, using a regularization parameter value of $1 \times 10^{-3}$.

\section{DSKO optimized parameters}
\begin{table}[!htp]
\caption{Woods-Saxon confinement parameters for Bi.}
\label{tab:bi_params_pbe}
\begin{tabular}{cccc}
\hline
Shell & $r_0$ & $a$ & $W$ \\
\hline
\hline
5d & 6.657 & 6.741 & 12.201  \\
6s & 3.610 & 7.605 & 42.474  \\
6p & 5.718 & 7.916 & 40.974  \\
\hline
Density & 9.453 & 7.183 & 33.927 \\
\hline
\end{tabular}
\end{table}

\begin{table}[!htp]
\caption{Spin constant matrix for spin-polarization calculations for Bi (in au).}
\label{tab:bi_spin_constants_pbe}
\begin{tabular}{c|ccc}
\hline
 & 5d & 6s & 6p \\
\hline
5d & -0.0135 & -0.00514 & -0.00183 \\
6s &-0.00514 & -0.0134 & -0.00946  \\
6p & -0.00183 & -0.00946 & -0.0109 \\
\hline
\end{tabular}
\end{table}

\section{Additional SOC band structure results} 

This section includes results for the SOC orbital projections of the rhombohedral phase band structure and the non-SOC band structure for the compressed rhombohedral phase.

\begin{figure}[!htp]
    \centering
    \subfloat[s-orbital.\label{fig:sorb}]{
        \includegraphics[
            width=0.45\columnwidth
        ]{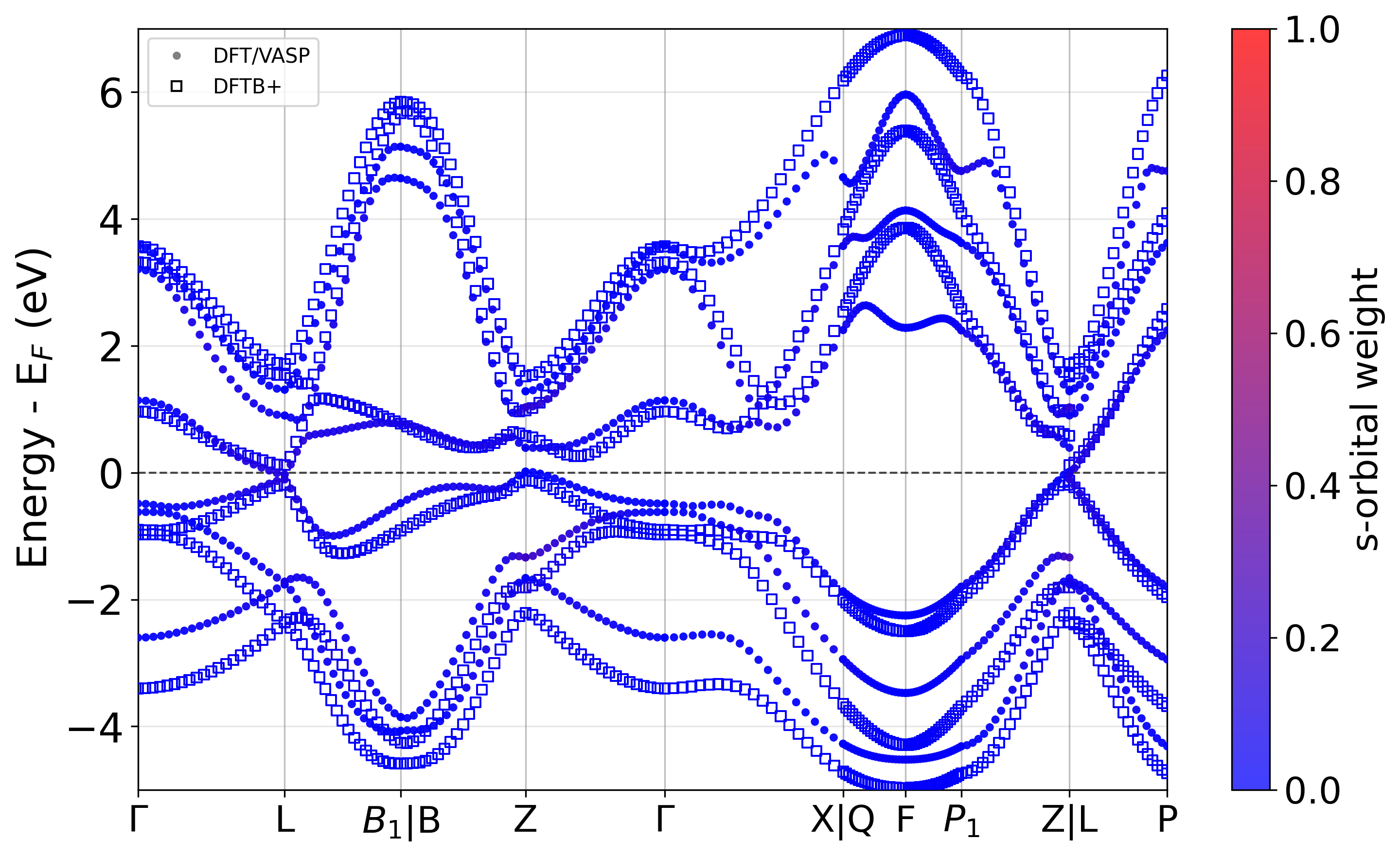}
    }
    \vspace{2mm}
    \subfloat[p-orbital.\label{fig:porb}]{
        \includegraphics[
            width=0.45\columnwidth
        ]{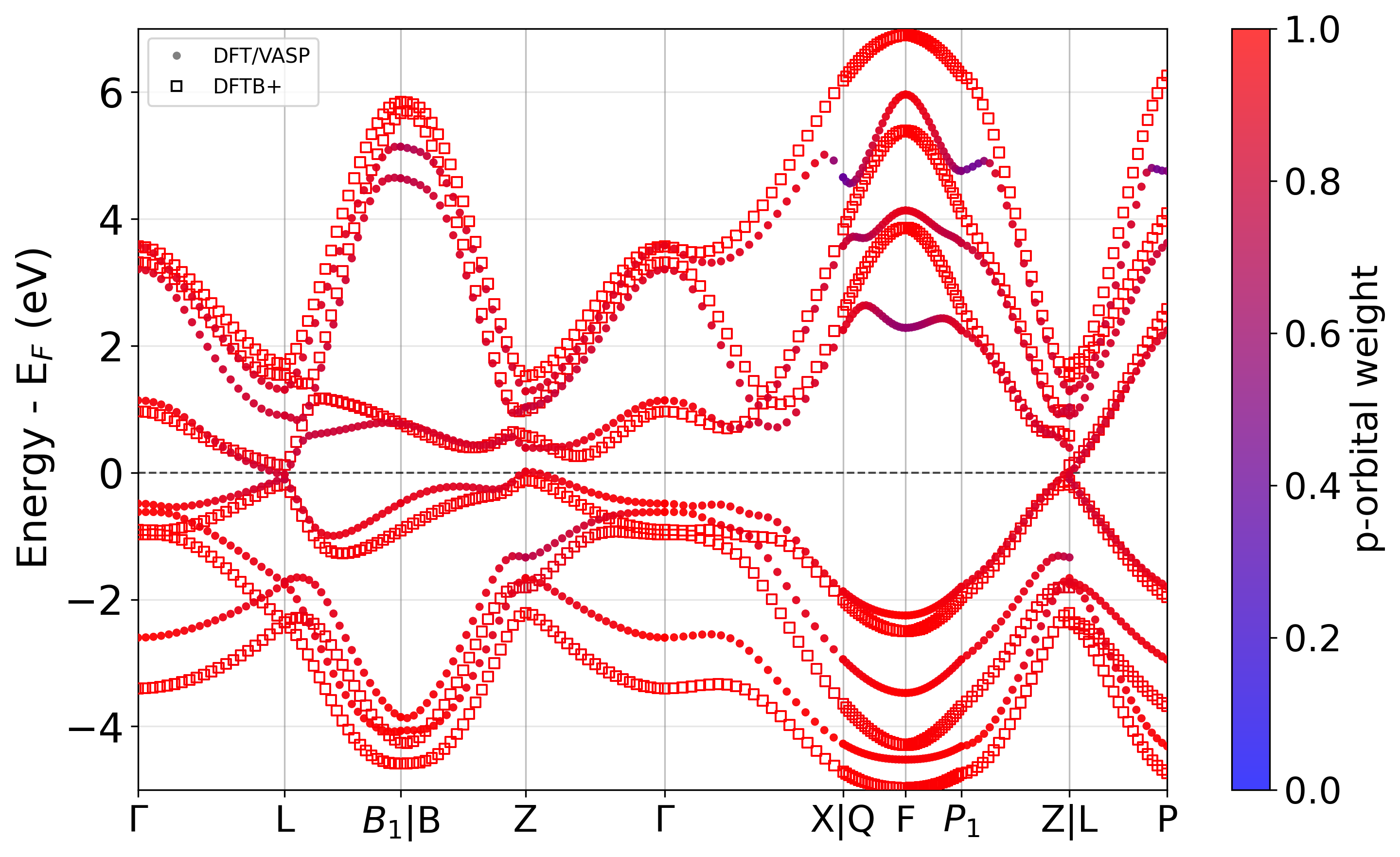}
    }
    \vspace{2mm}
    \subfloat[d-orbital.\label{fig:dorb}]{
        \includegraphics[
            width=0.45\columnwidth
        ]{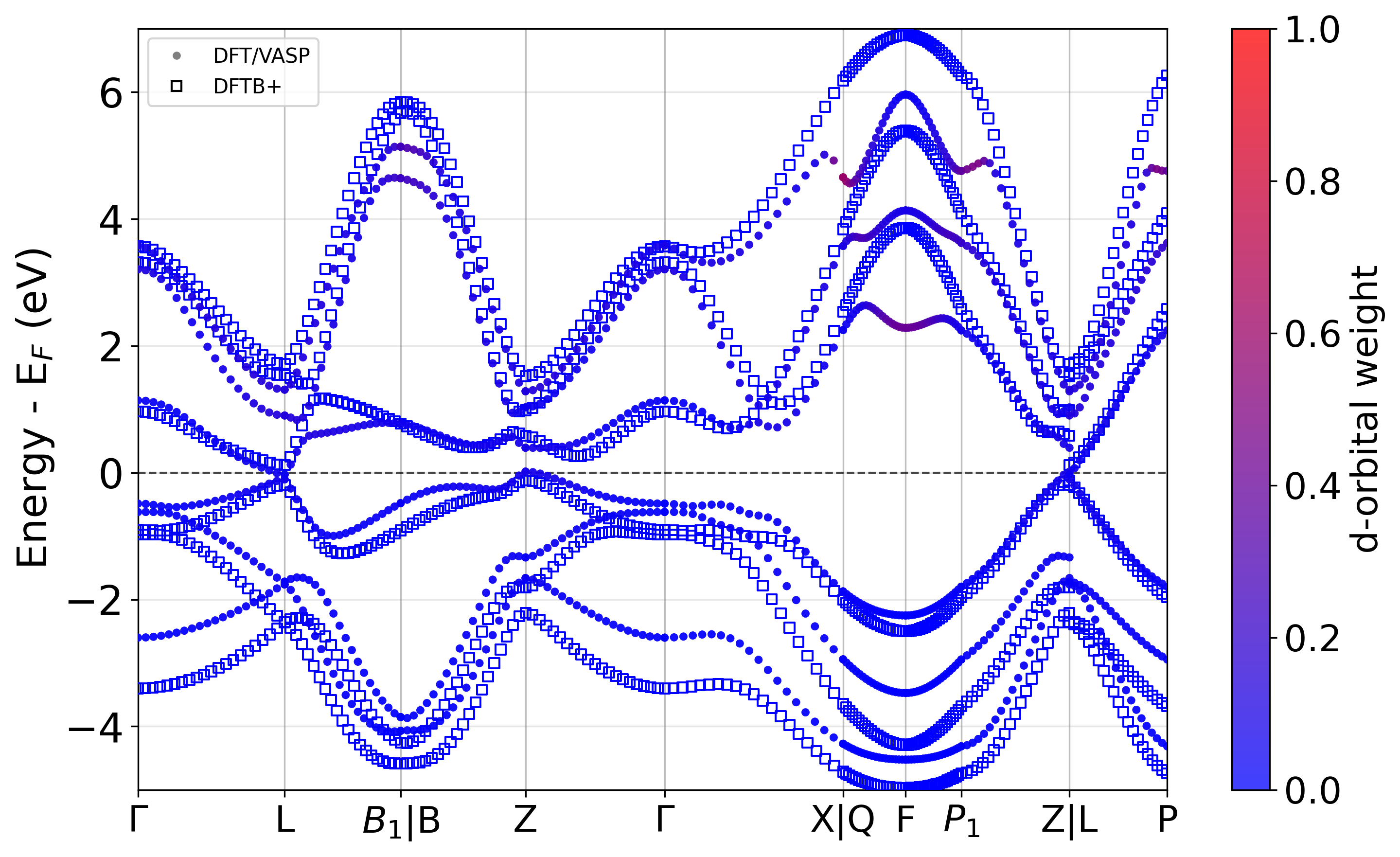}
    }
    \label{fig:contours}
    \caption{SOC orbital projections for uncompressed rhombohedral bismuth. Open squares correspond to DFTB and filled circles to DFT.}
\end{figure}

\begin{figure}[!htp]
    \centering
        \includegraphics[
            width=0.45\columnwidth
        ]{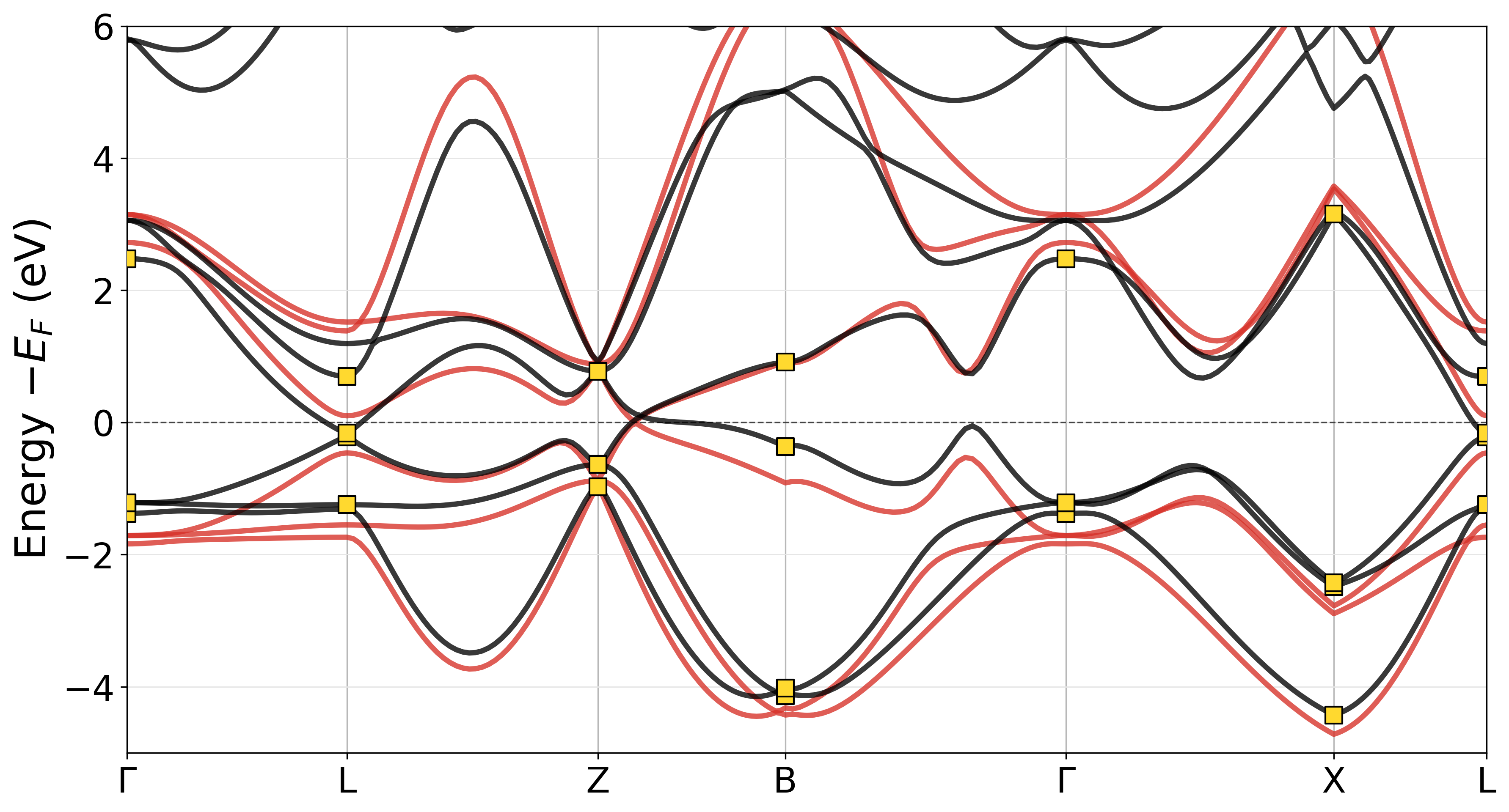}
    \caption{\label{fig:rhomb_compressed} Compressed rhombohedral phase band structure. Red bands correspond to DFTB/ChIMES and black to DFT.}
\end{figure}

\section{Bismuth crystal structure optimization results}

These are the DFTB/ChIMES results for the lattice optimizations for the rhombohedral, Bi-II, bcc, and simple cubic phases.

\begin{table*}[!htbp]
\centering
\caption{Relative energies ($\Delta E$), lattice parameters
($a$, $b$, and $c$), lattice angles ($\alpha$, $\beta$,
and $\gamma$), and nearest-neighbor distances (NN) for
the Bi polymorphs considered in this work.
Energies are reported in eV/atom relative to rhombohedral Bi.
Lengths are given in \AA\ and angles in degrees.}
\label{tab:bi_polymorphs}
\begin{tabular}{llcccccccc}
\hline
\hline
Structure & Method &
$\Delta E$ & $a$ & $b$ & $c$ &
$\alpha$ & $\beta$ & $\gamma$ & NN \\
\hline
Rhombohedral & DFT
& 0.00 & 4.88 & 4.88 & 4.88 & 56.7 & 56.7 & 56.7 & 3.12 \\
             & DFTB/ChIMES
& 0.00 & 4.79 & 4.79 & 4.79 & 58.7 & 58.7 & 58.7 & 3.20 \\
\hline
Bi-II        & DFT
& 0.06 & 6.94 & 6.49 & 3.47 & 90.0 & 116.7 & 90.0 & 3.27 \\
             & DFTB/ChIMES
& 0.04 & 7.13 & 6.54 & 3.37 & 90.0 & 111.7 & 90.0 & 3.33 \\
\hline
Simple cubic & DFT
& 0.05 & 3.30 & 3.30 & 3.30 & 90.0 & 90.0 & 90.0 & 3.30 \\
             & DFTB/ChIMES
& 0.05 & 3.33 & 3.33 & 3.33 & 90.0 & 90.0 & 90.0 & 3.33 \\
\hline
bcc          & DFT
& 0.22 & 4.03 & 4.03 & 4.03 & 90.0 & 90.0 & 90.0 & 3.49 \\
             & DFTB/ChIMES
& 0.22 & 4.09 & 4.09 & 4.09 & 90.0 & 90.0 & 90.0 & 3.54 \\
\hline
\hline
\end{tabular}
\end{table*}

\section{Calculation of the Fu-Kane inversion parity invariant}

The $\mathbb{Z}_2$ inversion symmetry invariant is determined for the occupied Kramers pairs at the TRIM points, namely $\Gamma$, $M_1$, $M_2$, and $M_3$. We first determine the inversion parity eigenvalues by constructing the spatial inversion operator $\mathbf{D}_\mathrm{spatial}$ with

\begin{equation}
\mathbf{D_\mathrm{spatial}} = \mathbf{B}(\mathbf{k})\mathbf{P}\mathbf{J}.
\end{equation}

\noindent Here, $B$ is the diagonal matrix that contains the Bloch phase factors due to the lattice translations associated with each atomic inversion,

\begin{equation}
\mathbf{B}(\mathbf{k}) =
\begin{pmatrix}
e^{-i\mathbf{k}\cdot\mathbf{T}_1} \mathbf{I}_9 & 0 \\
0 & e^{-i\mathbf{k}\cdot\mathbf{T}_2} \mathbf{I}_9.
\end{pmatrix}.
\end{equation}

\noindent The terms $\mathbf{T}_1$ and $\mathbf{T}_2$ correspond to the lattice translation vectors associated with the atomic inversion mapping and $\mathbf{I}_9$ to a $9 \times 9$ identity matrix. The resulting $\mathbf{B(k)}$ matrix has dimensions of $18 \times 18$ since the bismuthene primitive cell contains two atoms and each atom has nine valence orbitals (5d, 6s, 6p). The $\mathbf{P}$ matrix inverts the atomic positions and is similarly constructed from four $9 \times 9$ blocks,

\begin{equation}
\mathbf{P} =
\begin{pmatrix}
0 & \mathbf{I}_9 \\
\mathbf{I}_9  & 0.
\end{pmatrix}
\end{equation}

\noindent The matrix $\mathbf{J}$ is block diagonal and accounts for the intrinsic parity of the atomic orbitals under inversion, namely

\begin{equation}
(-1)^l =
\begin{cases}
+1 & \mathrm{s}\text{ orbitals }(l=0),\\
-1 & \mathrm{p}\text{ orbitals }(l=1),\\
+1 & \mathrm{d}\text{ orbitals }(l=2).
\end{cases}
\end{equation} 

We can then compute the spin-dependent inversion operator via $\mathbf{D}_{\mathrm{spin}} =
\mathbf{I}_2 \otimes \mathbf{D}_{\mathrm{spatial}}$ since the spin moments are not affected by spatial inversion. This yields an inversion matrix $\mathbf{M}_\mathrm{occ}$ that can be written as 

\begin{equation}
\mathbf{M}_{\mathrm{occ}} =
\mathbf{C}_{\mathrm{occ}}^\dagger
\mathbf{S}_{\mathrm{spin}}
\mathbf{D}_{\mathrm{spin}}
\mathbf{C}_{\mathrm{occ}}.
\end{equation}

\noindent In this case, $\mathbf{C}_{\mathrm{occ}}$ is the matrix of eigenvectors of the occupied states at a given TRIM point and $\mathbf{S}_{\mathrm{spin}}$ is the overlap matrix in the spinor atomic basis set, i.e., $\mathbf{S}_{\mathrm{spin}}(\mathbf{k}) =
\mathbf{I}_2 \otimes \mathbf{S}({\mathbf{k}})$. This yields dimensions of $n_\mathrm{occ} \times n_\mathrm{occ}$ for $\mathbf{M}_{\mathrm{occ}}$, where $n_\mathrm{occ}$ is the number of occupied states in the spinor representation. $\mathbf{M}_{\mathrm{occ}}$ can then be diagonalized to yield the parity eigenvalues, $\xi_n$. For a system that is inversion symmetric, $\xi_n=\pm1$, depending on the even/odd parity of the states. 

We can then compute the product of the eigenstates for each TRIM point, $\Lambda_i$, 

\begin{equation}
\delta_i = \prod_{m=1}^{N_{\mathrm{occ}}/2} \xi_{2m}(\Lambda_i).
\end{equation}

\noindent where we use only one parity eigenvalue from each occupied Kramers pair because time-reversal symmetry guarantees that the two states in a Kramers pair have the same inversion parity. The $\mathbb{Z}_2$ invariant is then computed from

\begin{equation}
(-1)^\nu =
\delta_\Gamma \delta_{M_1} \delta_{M_2} \delta_{M_3}.
\end{equation}

\noindent In this case, a value of $\nu = 0$ corresponds to trivial topology and $\nu = 1$ indicates a topological insulating system (assuming the system has a global band gap). 

\newpage
\bibliography{library}